\documentclass{article}
\usepackage{emulateapj}
\usepackage{epsfig}
\usepackage{flushrt}
\usepackage{apjfonts}

\newcommand{\bvfreq}{Brunt-V\"ais\"al\"a frequency}
\newcommand{\etal}{et al.}
\newcommand{\subs}[1]{_{\rm #1}}
\newcommand{\ylm}{Y_\ell^m(\theta,\phi)}
\newcommand{\Msun}{{M}_\odot}
\newcommand{\Mstar}{{ M}_\star}
\newcommand{\lhat}{\hat{\ell}}
\newcommand{\gras}{Geophys. J. Roy. Astr. Soc.}
\renewenvironment{figure}{\begin{figure*} }{\end{figure*}}

\begin{document}

\submitted{Received 1999 May 11; accepted 1999 July 2}

\title{The Effect of Crystallization on the Pulsations of
White Dwarf Stars}

\author{
M. H. Montgomery\altaffilmark{1,2} \&
D. E. Winget\altaffilmark{1} 
}

\altaffiltext{1}{McDonald Observatory and Department of Astronomy,
The University of Texas, Austin, TX 78712 -- USA.}
\altaffiltext{2}{Institut f\"ur Astronomie, Universit\"at Wien,
T\"urkenschanzstra\ss e 17, A-1180 Wien, Austria.}

\begin{abstract}

We consider the pulsational properties of white dwarf star models with
temperatures appropriate for the ZZ Ceti instability strip and with
masses large enough that they should be substantially crystallized.
Our work is motivated by the existence of a potentially crystallized DAV,
BPM 37093, and the expectation that digital surveys in  progress will
yield many more such massive pulsators.

A crystallized core makes possible a new class of oscillations, the
torsional modes, although we expect these modes to couple at
most weakly to any motions in the fluid and therefore to remain
unobservable. The $p$-modes should be affected at the level of a few
percent in period, but are unlikely to be present with observable
amplitudes in crystallizing white dwarfs any more than they are in
the other ZZ Ceti's.  Most relevant to the observed light variations
in white dwarfs are the $g$-modes. We find that the kinetic energy
of these modes is effectively excluded from the crystallized cores of
our models.  As increasing crystallization pushes these modes farther
out from the center, the mean period spacing $\langle \Delta P \rangle$
between radial overtones increases substantially with the crystallized
mass fraction, $M\subs{cr}/M_{\star}$. In addition, the degree and
structure of mode trapping is affected.  The fact that some periods are
strongly affected by changes in the crystallized mass fraction while
others are not suggests that we may be able to disentangle the effects
of crystallization from those due to different surface layer masses.

\end{abstract}


\section{Astrophysical Context}

The theoretical study of pulsating crystalline objects extends many years
into the past. One of the first numerical studies was by \cite{Alterman59}
(1959), who modeled global oscillations of the Earth. Their main interest
was in fitting the oscillation period of 57 minutes which was excited
by the Kamchatka earthquake of 1952.  In the process, they examined how
the central density in their models allowed them to match the periods of
other oscillation modes which were also observed to be excited by the
earthquake.

In an astrophysical context, \cite{Hansen79} (1979) treated oscillations
in white dwarf models with a crystalline inner core. Since it was
known that 1 $\Msun$ models with $T\subs{eff} \sim$ 10,000 K were in
the process of crystallizing (\cite{Lamb75} 1975; \cite{VanHorn76}
1976), \cite{Hansen79} self-consistently treated the response of the
crystalline core to the pulsations.  Their main interest was in explaining
the observed period ranges of the ZZ Ceti's in terms of low-radial order
oscillations. They found that the $g$-mode periods were decreased by the
presence of crystallization, contrary to our present findings.

\cite{McDermott88} (1988) treated oscillations in neutron star models
with a fluid core, a solid crust, and a thin surface fluid ``ocean.''
They considered neutron star oscillations as a possible explanation
for the observed irregularities in the timing of subpulses from radio
pulsars, and as a source of the observed periodicities in many X-ray
burst sources. They found $g$-modes which were trapped in the cores of
their models, as well as those which were trapped in the surface oceans.

Finally, \cite{Bildsten95} (1995) considered $g$-mode oscillations in
the thin surface oceans of accreting neutron star models. Their aim
was to explain the observed 5--7 Hz quasi-periodic oscillations in the
brightest accreting neutron star systems. They found a good match to
these frequencies for low order, $\ell=1$ $g$-modes.

Why, then, does this problem need to be re-examined in the context
of white dwarf stars? As is often the case, new observations and new
circumstances have again made this problem one worth considering, but
in more detail than the general analyses of the past. For example,
the pioneering calculations of \cite{Hansen79} (1979) were focussed
primarily on the range of normal mode periods which are possible given
a crystallized core, not with the details of how the periods of high
overtone $g$-modes are affected at the level of 5--10\%. At the time,
there were no known high-mass white dwarf pulsators, and precise
mode identifications for {\em any} pulsating white dwarf had yet to
be attempted.

\begin{figure}
{\hfill \epsfig{file=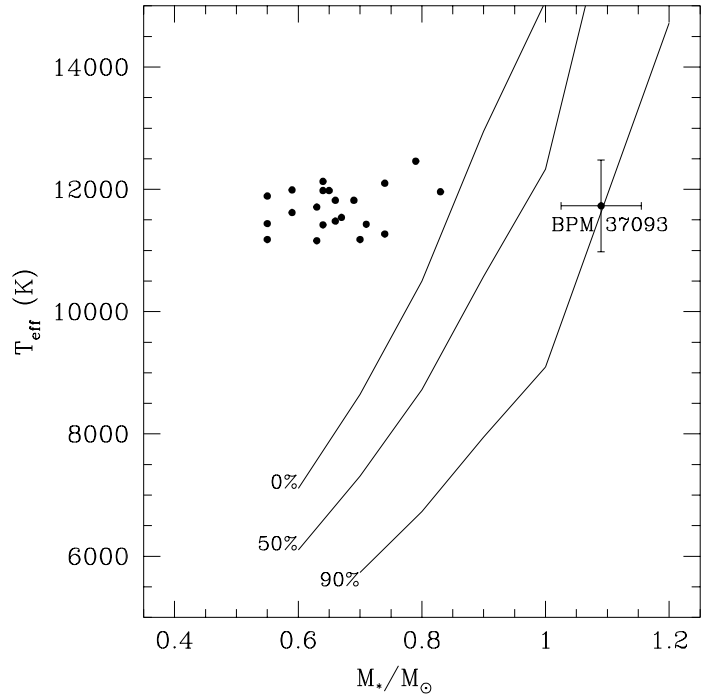,width=3.82in} \hfill}
\caption{The position of BPM 37093 relative to the other ZZ
Ceti's in Bergeron \etal\ 1995 as a function of $T\subs{eff}$ and
$M_{\star}/M_{\odot}$.  The lines correspond to constant amounts of
crystallized mass fraction assuming a pure oxygen core. If BPM 37093 has
an oxygen core it should be $\sim$ 90\% crystallized, and for a carbon
core, $\sim$ 50\% crystallized.
\label{bpm_pos}
}
\end{figure}

That situation changed with the discovery of pulsations in BPM
37093 (\cite{Kanaan92} 1992), a high-mass ZZ Ceti star (see Figure
\ref{bpm_pos}) which should be substantially crystallized (\cite{Winget97}
1997); depending on the C/O ratio in its core, it should be between
50\% and 90\% crystallized by mass. Depending on the details of its
nuclear history, its core could be composed of even heavier elements
such as Ne (\cite{Iben91} 1991), which would imply that it is more than
90\% crystallized (\cite{Winget97} 1997).  The Whole Earth Telescope
(WET) examined this target in the Spring of 1998 and found at least
8 independent frequencies, three of which had been previously seen by
\cite{Kanaan96} (1996). Thus, the potential to perform asteroseismology on
this object requires us to make a more detailed theoretical investigation
of the properties of crystallized pulsators.

One hope is that we will be able to independently determine the
crystallized mass fraction $M\subs{cr}/M_{\star}$, and thereby provide
a direct test of the theory of crystallization, now nearly four decades
old (\cite{Abrikosov60} 1960; \cite{Kirzhnits60} 1960; \cite{Salpeter61}
1961).  This subject is relevant to the astronomical community at large,
since phase separation of C and O during crystallization, and, indeed,
crystallization itself, represent the largest sources of systematic
uncertainties in the age of the local Galactic disk as derived from
the white dwarf luminosity function. In addition, understanding the
internal structure of white dwarfs may prove vital in fitting cosmological
models to Supernova Ia (SNIa) data (\cite{Garnavich98} 1998), so that
systematic differences in the absolute magnitudes of the SNIa may
be corrected for the evolutionary differences in the SN progenitors
(\cite{Hoeflich98} 1998).

Finally, digital surveys now in progress promise to add considerably to
the presently known number of cool white dwarfs. For instance, the Sloan
Digital Sky Survey (\cite{Gunn95} 1995) should increase this number by
a factor of approximately 20, with the result that we may have 20 such
stars with which to test the theory of crystallization.

\section{Evolutionary Models}

The basis for our equilibrium models is an updated version of WDEC, the
White Dwarf Evolutionary Code, as described in \cite{Lamb75} (1975)
and \cite{Wood90} (1990, \nocite{Wood92} 1992). Here we present only
a brief summary of the input physics in our models, with references
provided for a more complete description.

In the cores of our models we use the Lamb equation of state (EOS)
(\cite{Lamb74} 1974) and in the envelopes we use the tabular EOS of
\cite{Fontaine77} (1977). We employ the additive volume technique to
interpolate between pure compositions for the Carbon/Oxygen mixture in
the core and the Hydrogen/Helium/Carbon mixture in the envelope. The
chemical profiles of the composition transition zones in the envelope
are treated with an adaptation of the method of \cite{Arcoragi80}
(1980). Essentially, these profiles mimic those which would be obtained
in diffusive equilibrium, but contain additional parameters controlling
the thickness of the transition regions (\cite{Bradley93a} 1993).

The question of crystallization and our treatment of it is central
to our analysis. For a model with a given mass, $T\subs{eff}$,
and composition, the Lamb EOS does return a unique answer for the
degree of crystallization: the critical value of $\Gamma$ is given by
$\Gamma\subs{cr} \simeq 160$, where $\Gamma \equiv Z^2 e^2/ \langle r
\rangle k_B T$ is the ratio of Coulomb energy between neighboring ions to
each ion's kinetic energy. More recent calculations indicate a somewhat
higher value for this ratio, $\Gamma\subs{cr} \simeq 180$ (\cite{Ogata87}
1987). Our approach is to compute equilibrium models using WDEC and the
Lamb EOS, which results in models with a self-consistently computed value
of the crystallized mass fraction. When we perform a pulsational analysis
of these models, however, we take the crystallized mass fraction to be a
free parameter, in hopes of using asteroseismology to place constraints
on the degree of crystallization. The underlying assumption here is that
two equilibrium models which differ only in the degree of crystallization
have virtually identical pressure, density, and temperature profiles.

While this is not a physically self-consistent procedure, it is
justifiable for two reasons. First, the main physical effect of
crystallization is the release of latent heat; this provides the
models with an additional energy source, which means that at a
given $T\subs{eff}$ they are older.  This clearly has no effect
on the pulsational properties, which depend only on the structural
parameters of a given equilibrium model. Second, the density change
at crystallization is quite small, $\delta \rho/\rho \sim 10^{-3}$
(\cite{Lamb75} 1975), so the difference in, for example, $\rho(r)$,
$P(r)$, and $T(r)$ between two models which differ only in the amount
of crystallization is accordingly quite small. As we show in section
\ref{hardsphere}, the effect of crystallization upon $g$-mode pulsations
can be accurately taken into account through a modified boundary condition
at the assumed solid/fluid interface.

\section{Asymptotic Nonradial Oscillation Theory}

Stars which are fluid (uncrystallized) can undergo nonradial motions
which have been labelled $g$- and $p$-mode oscillations. In the linear limit,
these modes of oscillation are spheroidal, with the Eulerian perturbations
of variables such as the density and pressure having the angular spatial 
dependence of a single spherical harmonic (i.e., $\rho^{\prime}$, $p^{\prime}
\propto \ylm$). From a local analysis, the radial wavenumber $k_r$ is given by
\begin{equation}
k_r^2 = \frac{1}{\sigma^2 c_s^2} (\sigma^2 - L_{\ell}^2) (\sigma^2-N^2),
\end{equation}
where $\sigma$ is the angular frequency of the mode, $c_s$ is the sound
speed, $L_{\ell}^2 \equiv \ell (\ell+1) c_s^2/r^2$ is the square of
the Lamb/acoustic frequency, $r$ is the radial variable, and $N^2$
is the famed \bvfreq\ (see \cite{Unno89} 1989 for a more complete
discussion). From the above formula, we see that a mode is propagating
in a region (i.e., has $k_r^2 \geq 0$) if $\sigma^2 > L_{\ell}^2, N^2$,
{\em or} $\sigma^2 < L_{\ell}^2, N^2$. Thus, the modes separate cleanly
into two classes:

\begin{description}
\item[$p$-modes] \mbox{} \\
   $\sigma^2 > L_{\ell}^2, N^2$, ``high-frequency limit''\\
   $\sigma_k \sim \frac{k \pi}{\int_{r1}^{r2} dr/c_s}$ \\
   displacements become vertical near the surface
\item[$g$-modes] \mbox{} \\
   $\sigma^2 < L_{\ell}^2, N^2$, ``low-frequency limit''\\
   $P_k \sim \frac{2 \pi^2 k}{\sqrt {\ell (\ell+1)} } 
   \left[\int_{r1}^{r2} N dr/r \right]^{-1}$ \\
   displacements become horizontal near the surface
\end{description}
Here $r1$ and $r2$ are the inner and outer classical turning points,
respectively, at which $k_r = 0$ for a given $\sigma$.
We see that in the asymptotic limit the $p$-modes are uniformly spaced
in frequency as a function of radial order $k$, while the $g$-modes
are uniformly spaced in period. 

\begin{figure}
{\hfill \epsfig{file=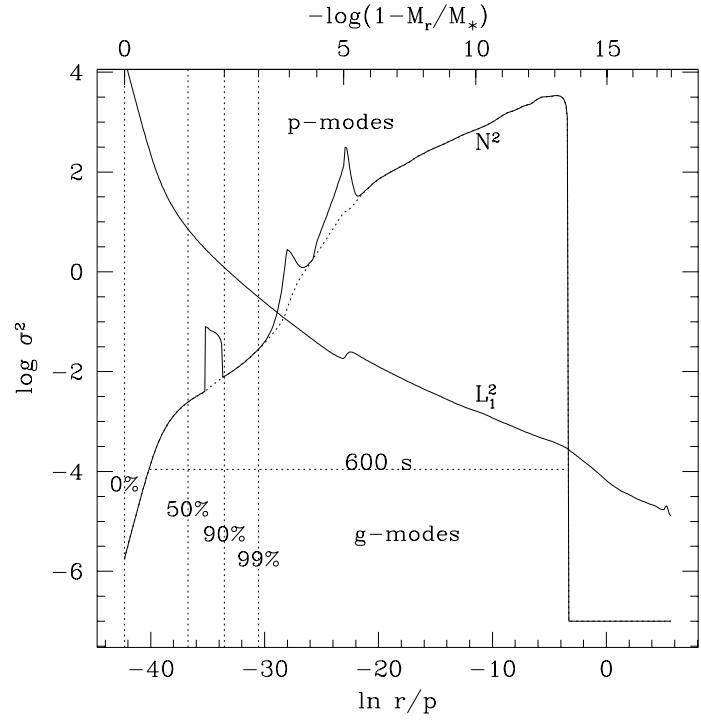,width=3.82in} \hfill}
\caption{A propagation diagram showing $N^2$ and $L_{1}^2$ as a function
of $\ln \,(r/p)$ (lower axis) and $-\log (1-M_r/M_{\star})$ (upper axis); the
center is on the left and the surface is on the right. The region of
propagation of a 600 second $g$-mode is shown. The vertical dashed lines
are labelled by the percent mass which is interior to these regions, i.e.,
the 90\% line indicates the boundary at which 90\% of the mass of the model
is inside this point. We see that a model which is this crystallized now
has an inner turning point for $g$-mode propagation considerably farther
out than in the uncrystallized case.
\label{prop}
}
\end{figure}

A useful diagnostic for the frequency spectrum of a given white dwarf
model is the propagation diagram, an example of which is shown in Figure
\ref{prop}, where we have labelled the high and low frequency domains
of the $p$- and $g$-modes; the model is of a 1.1 $\Msun$ white dwarf
with $T\subs{eff} = $ 12,200 K.  The horizontal axis is given in terms
of the radial variable $\ln \,(r/p)$, which is the natural logarithm of the
radius divided by the pressure, where both $r$ and $p$ are given in cgs
units. This radial variable has the desirable property that it increases
monotonically outward from the center and provides increased resolution
in both the core (where $r$ approaches zero) and in the envelope (where $p$
approaches zero). Along the upper axis we display the more commonly used
radial variable $-\log (1-M_r/M_{\star})$, which may be more easily
related to the structural parameters of the models.

We note that the bumps in $N^2$ and $L_{\ell}^2$ correspond to the
C/O, He/C, and H/He transition zones. For instance, using the upper
axis to obtain estimates of $-\log (1-M_r/M_{\star})$, we see that
the outer two transition zones have $M\subs{H}/M_{\star} \sim 10^{-5}$,
$M\subs{He}/M_{\star} \sim 10^{-3}$, which are in fact the values assumed
in these models.

\section{The Effect of a Crystalline Core}

How does a crystalline core affect the oscillations of a star? As we
introduce a solid core into our models, two things occur: (1) A new class
of modes appears (the torsional/toroidal modes, in this case), and (2)
the pre-existing $p$- and $g$-modes are modified.  We now treat these
cases separately.

\subsection{The Torsional Modes}

The torsional modes, or $t$-modes, are very special nonradial modes
characterized by zero radial displacement and zero compression, i.e.,
$\xi_r$ and $\vec{\nabla}\cdot \vec{\xi}$ both vanish, where $\vec{\xi}$
is the displacement vector. The dispersion relation for these modes is
\begin{equation}
  k_r^2=\frac{1}{v_s^2} (\sigma^2-T_{\ell}^2),
\end{equation}
where $v_s^2 = \mu/\rho$ is the square of the shear velocity, $\mu$ is the
shear modulus, and $T_l^2 = [\ell (\ell+1)-2] v_s^2$ is the ``torsional
frequency.'' They propagate in the region defined by $\sigma^2 > T_{\ell}^2$, 
and their frequency spectrum is equally spaced, as is the case
with $p$-modes, with 
\[ \sigma_k \sim \frac{k \pi}{\int dr/v_s}. \] 
As we might expect, the $k=1$ period for these modes goes like 
$R_{\star}/v_s$, the crossing time for a shear wave. 

\begin{figure}
{\hfill \epsfig{file=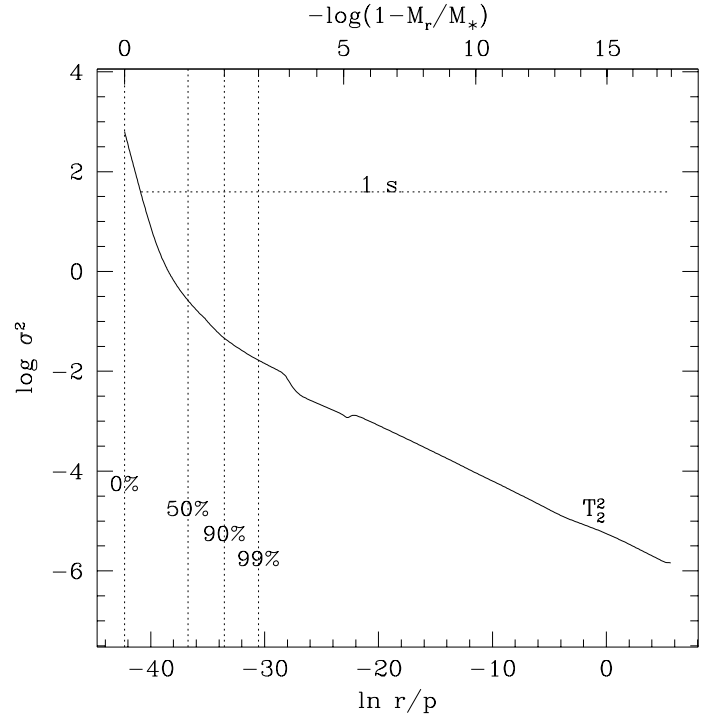,width=3.82in} \hfill}
\caption{A propagation diagram for $t$-modes for $\ell =2$. The $t$-modes
propagate only in the crystallized region, e.g., only to the left of the
90\% crystallized line for a 90\% crystallized model. We note that for
$\ell=1$ we have $T_{1}^2=0$, so the modes propagate throughout the entire
crystallized region in this case.
\label{tprop}
}
\end{figure}

In Figure \ref{tprop}, we show a propagation diagram for $t$-modes with
$\ell=2$, using the same white dwarf model as in Figure \ref{prop}.
If we imagine a model which is 90\% crystallized, then the $t$-mode
can potentially propagate anywhere inside the 90\% mass point in the
model. If the mode is an $\ell=2$ mode, then its region of propagation
is restricted further to the region for which its frequency is greater
than the torsional frequency, i.e., $\sigma^2 > T_{2}^2$. For a 1
sec mode, this corresponds to the part of the horizontal dotted line
which lies to the left of the 90\% point in Figure \ref{tprop}. We note
that for all the $\ell=1$ modes, we have $T_{1}^2=0$, so these modes
propagate throughout the entire crystallized region.

\setcounter{footnote}{0}

Observable consequences of the $t$-modes, if any exist, are difficult to
identify. The longest period $t$-modes should have periods $\sim$ 20 sec,
which is too short to explain the observed oscillations in the ZZ Ceti's
of 100's of seconds.\footnote{Such short timescale oscillations of 10's
of seconds may be relevant for accreting white dwarfs in cataclysmic
variable systems, although the heating due to accretion may preclude the
presence of substantial crystallization in these objects.} In addition,
these modes should not be able to couple (in the linear limit) to the
fluid at the solid/fluid interface, so these oscillations should be
unable to propagate from the crystalline core through the fluid to the
surface. Also, the different angular structure of the $t$-modes should
make any nonlinear coupling between these modes and the ordinary $p$-
and $g$-modes very weak; to the first nonlinear order this coupling will
be zero.  As a result, we expect these modes to be unobservable unless
crystallization has proceeded out into the photosphere. The oldest
known white dwarfs in the Galaxy are not yet cool enough for this to
have occurred. We therefore turn our attention to the $p$- and $g$-modes.

\subsection{The Spheroidal Modes}
\subsubsection{$p$-modes}
For pressure waves traveling in a solid medium, the velocity
$v_p$ is given by
\[ v_p^2 = \frac{\lambda + 2 \mu}{\rho} \]
where $\lambda = \Gamma_1 P -2 \mu/3$, $\rho$ and $P$ are the density and
pressure, respectively, and $\Gamma_1$ is the usual adiabatic exponent
(\cite{Landau75} 1975). If we treat the non-zero $\mu$ as a perturbation,
we find that
\[ \frac{\delta v_p}{v_p} \sim \frac{2 \mu}{3 \Gamma_1 P}, \]
where $\delta v_p$ is the change in $v_p$ due to the finite shear
modulus. In the cores of our 1.1 $\Msun$ models, we typically find $\mu/p
\sim 0.01$. Thus, $p$-mode periods are affected at the level of only a
few percent by the presence of a crystalline lattice. They are therefore
of no more interest than are ordinary $p$-modes in the context of the
observed pulsations of the DAV and DBV white dwarfs.

\subsubsection{$g$-modes}
We concentrate the remainder of our analysis on the $g$-modes, since
these are the modes which are believed to be responsible for the
observed pulsations in the white dwarf variables. Because $g$-modes
have large shears associated with their fluid motions, we expect
the nonzero shear modulus $\mu$ of the solid to have a significant
effect on them. Qualitatively, we may ask when the return force due
to a finite shear modulus is approximately equal to the return force
normally experienced by fluid elements in the absence of such shear
(e.g., \cite{Bildsten95} 1995). Algebraically, the shear return force
is equal to or exceeds the ordinary return force of the fluid when
\[ \frac{\mu}{\rho \sigma^2 h^2} \geq 1, \] 
where $h \equiv P/|dP/dr|$ is a pressure scale height.

In our models, we find that $\frac{\mu}{\rho \sigma^2 h^2} > 10^{10}$,
which indicates that the $g$-modes are completely altered in the crystallized
region. Thus, a $g$-mode which is propagating in the fluid region will
find a complete mismatch as it attempts to propagate into the crystallized
region. We therefore expect nearly complete reflection of the $g$-mode
at such a boundary, with the result that the $g$-modes are essentially
confined to the fluid regions of our models.

\section{Numerical Analysis}
\subsection{The Global Solution}
We now examine the above assertion and offer a numerical justification for
it. Our approach is based on the work of \cite{Hansen79} (1979); we treat
the ``global'' problem in that we allow the solid cores of our models to
respond to the oscillations.  We have used the Cowling approximation to
simplify the pulsation equations, as was also done in \cite{Hansen79}
(1979). Since $g$-modes in white dwarfs are primarily envelope modes,
this is an excellent approximation and hardly affects the accuracy of
our calculated periods; even $k=1$, $\ell=2$ modes have periods which
are only affected at the level of 0.2\% (\cite{Montgomery98a} 1998).
The details of the rest of the global treatment are summarized in
Appendix \ref{derivation}, where we describe the oscillation variables,
the equations which they obey, the central boundary conditions, and the
connecting conditions at the solid/fluid interface.

\begin{figure}
{\hfill \epsfig{file=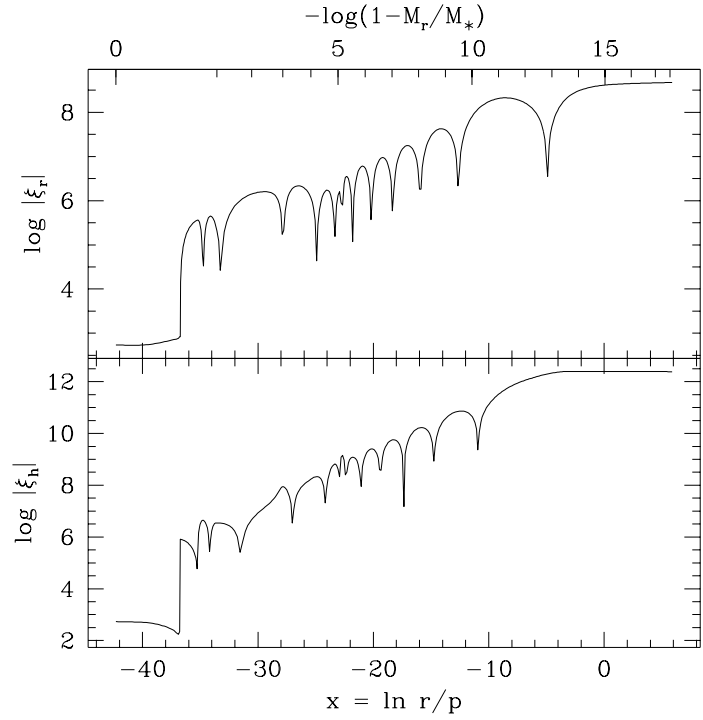,width=3.82in} \hfill}
\caption{The log of the absolute values of the radial (upper panel) and
horizontal (lower panel) displacements as a function of $\ln \,(r/p)$.
Note that $\xi_r$ is continuous at the solid/crystal interface at $\ln
\,(r/p) \sim -36.7$, but that $\xi_h$ is not. The magnitudes of both
$\xi_r$ and $\xi_h$ are reduced by $\sim$ 3 orders of magnitude as they
penetrate the solid region.
\label{xeigen} }
\end{figure}

In Figure \ref{xeigen}, we plot the radial and horizontal displacements of
a 378.4 sec, $\ell$ = 1 mode; the model is again that of a 1.1 $\Msun$
white dwarf with $T\subs{eff}$ = 12,200 K, which is assumed to be
50\% crystallized. As is true of all the $g$-modes we have examined,
the amplitude of the fluid motions is decreased by $\sim$ 3 orders of
magnitude in the solid as compared to the fluid. One other feature of
the oscillations is that the horizontal displacement is discontinuous
at the solid/liquid interface. In the approximation of zero viscosity
and laminar flow, the fluid is free to slide over the solid surface. In
reality, a turbulent boundary layer would probably form in this region,
which would tend to dissipate the pulsation energy.

The kinetic energy density depends on the square of the displacement,
so it is attenuated by $\sim 6$ orders of magnitude in the solid core.
Since the kinetic energy is an indicator of how a given mode samples
the different regions of a model, we conclude that it is a very good
approximation to treat the $g$-modes as excluded from the solid cores
of our models. In the following section we will demonstrate the validity
and the self-consistency of this approach.

\subsection{The ``Hard-Sphere'' Boundary Condition}
\label{hardsphere}
As suggested in the previous section, we may be able to reproduce the
effects of crystallization on $g$-mode pulsations merely by applying a
hard-sphere boundary condition at the solid/liquid interface. By this
we mean that the radial displacement is set to zero ($\xi_r=0$) and the
horizontal displacement is left to be arbitrary. In addition, the boundary
condition on the gravitational potential and its derivative are the same
as for the uncrystallized case, as we show in appendix \ref{boundary};
these ``hard-sphere'' calculations are not in Cowling approximation and
therefore solve the full fourth-order adiabatic equations in the fluid
region.  Using the hard-sphere boundary condition has the advantage that
the resulting problem is much easier to treat, both in terms of speed
and convergence.

We have calculated the fractional difference between periods calculated
with the ``hard-sphere'' approximation and those calculated with the
``global'' treatment. Using a fiducial model with $\Mstar = 1.1 \Msun$,
$T\subs{eff} = 12,200$ K, and assuming 90\% crystallization by mass, we
have examined all $\ell=1$ and 2 periods between 50 and 1000 sec.  We find
that the fractional difference in periods is less than 1 part in $10^4$,
and that the absolute error in the calculated periods never exceeds 0.05
sec. We therefore conclude that the ``hard sphere'' boundary condition at
the solid/fluid interface accurately represents the physics of $g$-mode
oscillations in models with crystalline cores. \cite{Bildsten95} (1995)
found exactly the same approximation to be valid in their treatment of
$g$-modes in the surface oceans of accreting neutron star models.

\begin{figure}
{\hfill \epsfig{file=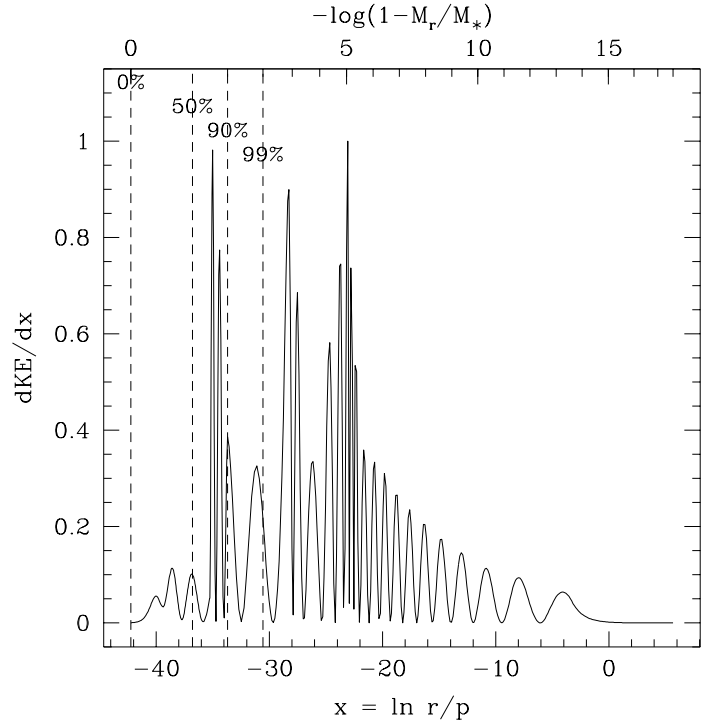,width=3.82in} \hfill}
\caption{The kinetic energy per unit $x = \ln \,(r/p)$. The vertical dashed
lines are labelled with the mass-fraction of the model interior to the
given point. The mode shown is an $\ell=1, k=25$ mode with a period of
673.4 sec.  The equilibrium model is a 1.1 $\Msun$ model with $T\subs{eff}$
at 12,200 K, and the surface layer masses are $M\subs{He}/\Mstar = 10^{-3}$
and $M\subs{H}/\Mstar =10^{-5}$.
\label{kinden}
}
\end{figure}

Before proceeding to the detailed numerical calculations, we wish to
convince the reader that crystallization will have a measurable effect on
the periods.  In Figure \ref{kinden} we have plotted the kinetic energy
per unit $x = \ln \,(r/p)$, so that the area underneath the curve represents
the weight of each region's contribution to the total kinetic energy
as a function of $x$. The vertical dashed lines indicate different mass
points in this model. For instance, if the model is 90\% crystallized,
then the kinetic energy to the left of the 90\% line is eliminated from
the mode. By visual inspection, this is of order 10\% of the kinetic
energy in the mode, so we might well expect that the period of this mode
is affected at the 10\% level.  In fact, we will see in the next section
that the periods can be shifted by even larger amounts.

\section{The $g$-mode Periods as a Function of $M\subs{cr}/\Mstar$}
\subsection{Asymptotic Relations}

The kinetic energy argument in section \ref{hardsphere} leads us to
expect that the $g$-mode periods will change measurably as the crystallized
mass-fraction increases from 0 to 90\%. With this in mind, we re-examine
the asymptotic formulae for $g$-mode periods (e.g., \cite{Unno89} 1989):
\[ P_k \sim k \langle \Delta P \rangle, \]
\begin{equation}
\langle \Delta P \rangle = \frac{2 \pi^2}{\sqrt {\ell (\ell+1)} }
   \left[\int_{r1}^{r2} N dr/r \right]^{-1},
\label{asympx}
\end{equation}
where we have written $\langle \Delta P \rangle$ for the mean period
spacing between consecutive radial orders. Since the $g$-modes are
excluded from the crystallized region, the inner turning point $r1$
is now a function of $M\subs{cr}/M_{\star}$.  As we allow the model
to crystallize while holding all other structural parameters constant,
$r1$ moves outward, so the integral in equation \ref{asympx} decreases,
with the result that $\langle \Delta P \rangle$ and $P_k$ both increase.

As a heuristic tool, we would like to plot the ``region of period
formation,'' which would tell us visually the weight which the different
regions of the star have in determining the period of a mode. This
problem has been examined several times in the past, for example by
\cite{Kawaler85a} (1985), \cite{Schwank76} (1976), \cite{Goosens74}
(1974), and originally by \cite{Epstein50} (1950).  

To simplify matters, we examine this weight function in the asymptotic
limit of high $k$ and $\ell$.  For $g$-modes, we have to be content to
determine a ``region of frequency formation.'' From asymptotic theory, 
we find that the relative contribution to the total frequency per
unit radius is
\[ \frac{d \sigma}{dr} \approx \frac{N}{r}, \]
which depends only on $N$ and $r$. In order to expand the radial axis
in both the center and the envelope and to make the resulting functions
easier to examine, we choose $x= \ln \,(r/p)$ as our radial coordinate. Then
the above relation becomes
\begin{equation} \frac{d \sigma}{dx} \approx \frac{N}{1+V},
\label{pformeqn1}
\end{equation}
where $V \equiv \Gamma_1 g r/c_s^2$. We emphasize that the appearance
of the sound speed $c_s^2$ in the variable $V$ is purely a result of
the above radial coordinate change, and does not reflect a dependence
of $g$-mode frequencies on $c_s$.

\begin{figure}
{\hfill \epsfig{file=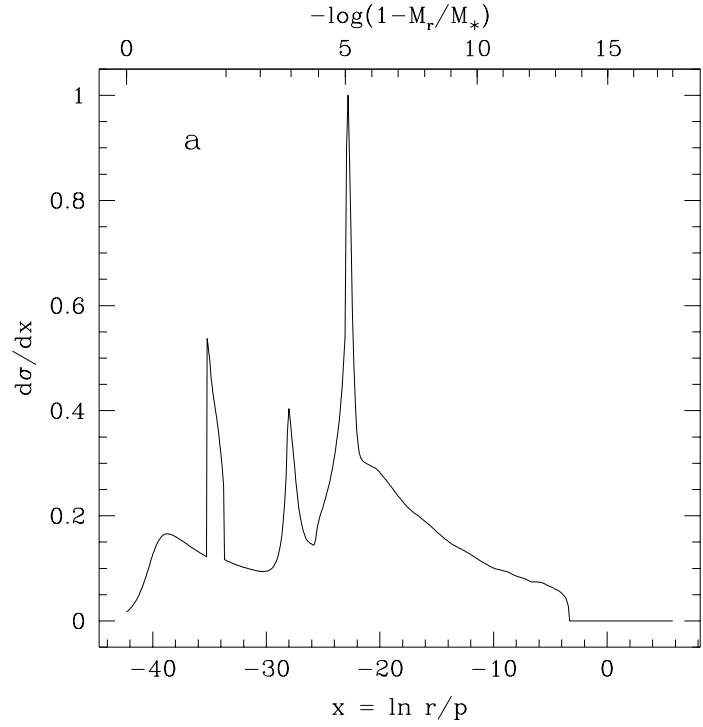,width=3.82in} \hfill}
\caption{The frequency (period) formation region for $g$-modes in a 1.1
$\Msun$ model with $T\subs{eff} = 12,200$~K. The three spikes are all
composition transition zone features, which from left to right are due
to the O/C, C/He, and He/H transition zones.
\label{pformg}
}
\end{figure}

In Figure~\ref{pformg}, we plot $d \sigma/dx$ versus $x$ for a 1.1
$\Msun$, $T\subs{eff}$ = 12,200 K model with $M\subs{He}/\Mstar =
10^{-3}$ and $M\subs{H}/\Mstar =10^{-5}$. The three spikes in $d\sigma/dx$
correspond to the composition transition zones of O/C, C/He, and
He/H. From inspection of this figure, we would expect the C/He transition
zone to have the least effect on the $g$-mode periods, whereas the
He/H transition zone in the envelope should have the largest effect.
Numerically, \cite{Bradley93b} (1993) has found this to be the case,
with the period spacing and mode trapping being most sensitive to the
hydrogen layer mass and least sensitive to the thickness of the helium
layer. Physically, this is due to the fact that the He/H transition zone,
since it is closest to the surface, is the least degenerate, so it has
the largest thermal contributions to the \bvfreq. In addition, it is the
only zone in which there is a contrast in the atomic weight per electron,
$\mu_e$. In going from He to H, $\mu_e$ goes from 2 to 1; for the O/C and
C/He zones, $\mu_e = 2$ for both chemical species in the transition zone.

It is worth commenting on the similarities between the distribution of
kinetic energy in Figure~\ref{kinden} and the shape of the $g$-mode
period formation region in Figure~\ref{pformg}. The kinetic energy
plot is for a numerically calculated $\ell=1, k=25$ mode, whereas
the period formation region is in the high $k$ limit. The value of
$N/(1+V)$ in Figure~\ref{pformg} should correspond to to the wavelength
of oscillations as a function of $x$ in Figure~\ref{kinden}.  This is in
fact the case, since we see that peaks in Figure~\ref{pformg} correspond
to rapid spatial oscillations in the kinetic energy density.  Similarly,
the small value of $N/(1+V)$ in Figure~\ref{pformg} for $x$ in the range
of $-10$ to $-4$ results in a longer spatial wavelength in the oscillations
of the kinetic energy density in Figure~\ref{kinden} at this value of $x$.

We also note that the overall envelope of the kinetic energy is similar
in shape to Figure~\ref{pformg}. With this in mind, we calculate the
kinetic energy distribution in the asymptotic limit. The kinetic energy
$dE$ in a shell $dr$ is given by
\begin{eqnarray*}
dE & \approx & \rho r^2 dr \left[\xi_r^2 + \ell(\ell+1) \xi_h^2\right] \\
  & \approx & \rho r^2 dr \xi_h^2,
\end{eqnarray*}
where we have used the fact that $\xi_h \gg \xi_r$ for $g$-modes. If we now
substitute for $\xi_h$ the asymptotic value for it taken from \cite{Unno89}
(1989), then we find that the envelope of the kinetic energy density
varies like
\begin{eqnarray}
\frac{dE}{dr}  & \approx & \rho r^2 \xi_h^2 \nonumber \\
 & \approx & \frac{N}{r}.
\end{eqnarray}
Thus, we see that in the asymptotic limit the kinetic energy samples
the model in the same way as does the frequency for $g$-modes.

Although the above discussion might give the impression that modes of
this radial overtone number are safely in the asymptotic limit, such is
not the case. For the mode in Figure~\ref{kinden}, it may be treated in
the asymptotic limit in the region between $x=-20$ and $x=-5$, i.e.,
the variations of its amplitude are small compared to its wavelength.
However, the composition transition zone at $x \sim -23$ provides a much
more rapid spatial variation than the wavelength of this mode. 
Depending on the details of how the mode interacts with this feature, it
will be partially transmitted and partially reflected at this boundary.
Thus, the amplitudes of the mode on each side of a transition zone
will not in general be given by the asymptotic theory. In other words,
the effect of a transition zone is to enhance the amplitude of a mode
on one side  of a transition zone relative to its amplitude on the
other side.  This effect is generically known as ``mode trapping,''
although in the context of white dwarfs this term usually denotes a
mode which has an enhanced amplitude in the outer surface layer, i.e.,
the H layer for DAV's. Returning to the mode in Figure~\ref{kinden},
neighboring modes which differ from it by only $\pm1$ in $k$ still have
somewhat different distributions of kinetic energy between the different
transition zones.  Thus, we cannot consider these modes to be globally
described by asymptotic theory.

\subsection{Numerical Results}

\begin{figure}
\plottwo{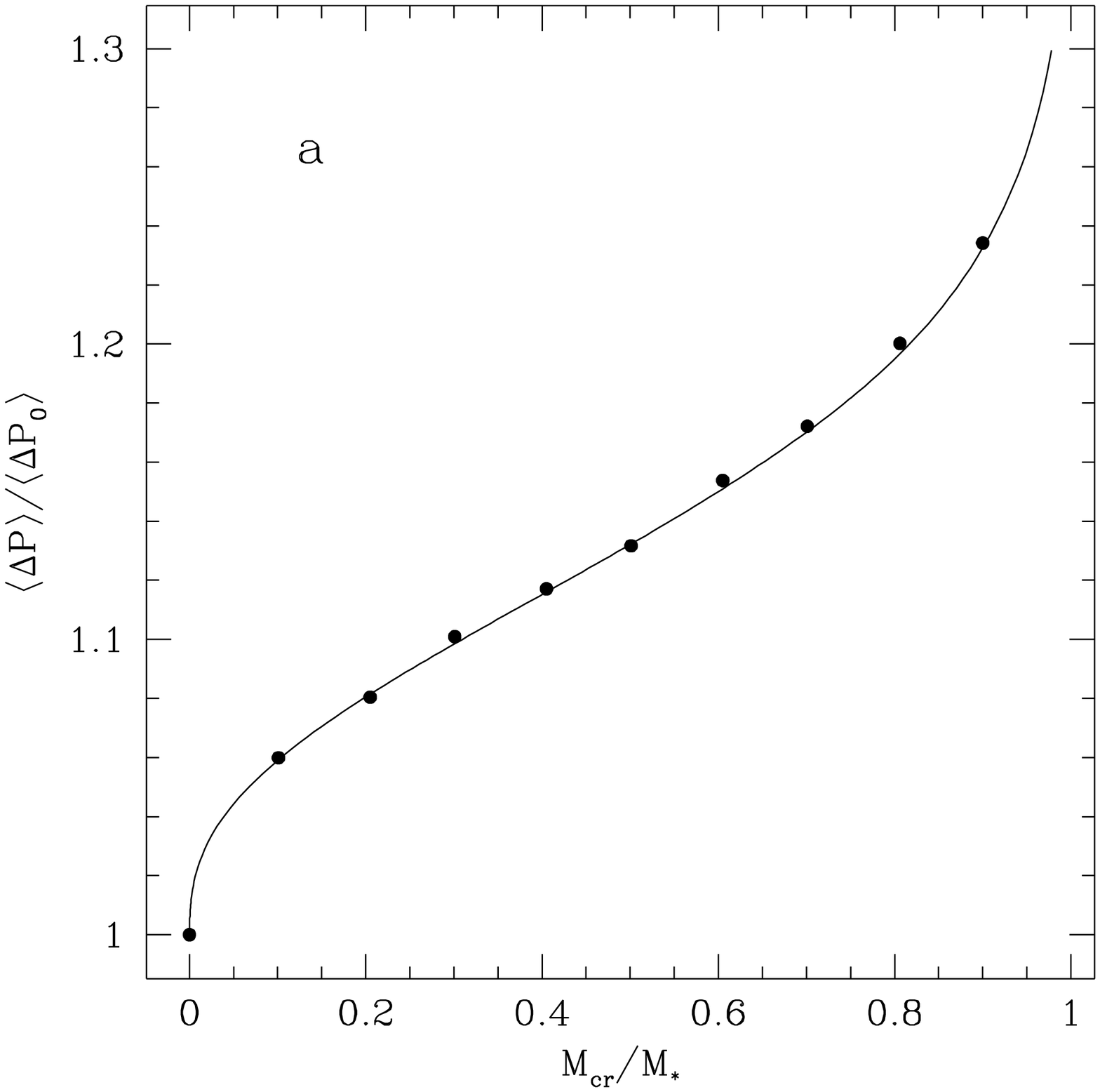}{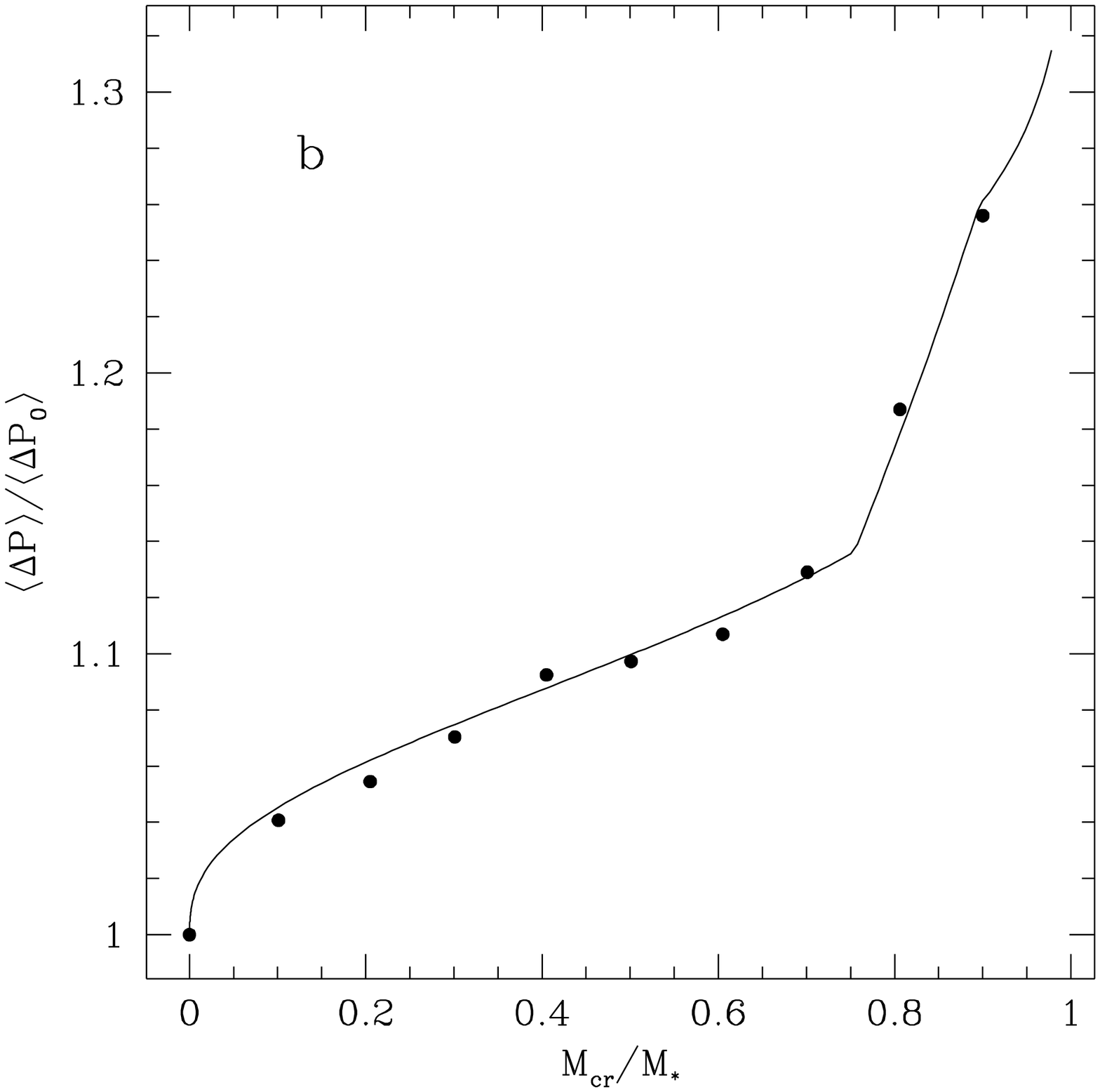}
\caption{A comparison of analytical (solid line) and numerical (filled
circles) period spacings, as a function of $M\subs{cr}/\Mstar$, where
each has been normalized to the period spacing in the uncrystallized
case. In order to minimize mode trapping effects, the Schwarzschild
criterion has been used to compute the \bvfreq\ in (a), whereas in
(b) the modified Ledoux prescription has been used. The ``kink'' in (b)
for $0.75 \leq M\subs{cr}/\Mstar \leq 0.90$ is due to the changing
C/O profile in the core.
\label{dp_x_schwz}
}
\end{figure}

We now wish to make a comparison between the functional form of the
period spacing implied by equation \ref{asympx} and that derived from
direct numerical calculations. To do this, we normalize $\langle \Delta P
\rangle$ to the average period spacing in the uncrystallized case, denoted
by $\langle \Delta P \rangle_0$. Such a comparison is shown in Figure
\ref{dp_x_schwz}a, where the solid line gives the analytic relation and
the filled circles are the result of a numerical pulsational analysis of
$\ell=2$ periods between 500 and 1000 sec. We have made the model, a 1.1
$\Msun$ C/O core model with $T\subs{eff}=12,200$ K, artificially smooth
by using the Schwarzschild criterion for the \bvfreq, which essentially
removes the bumps from the \bvfreq\ and therefore minimizes mode trapping.
The agreement between the two methods is extremely good.

\begin{figure}
{\hfill \epsfig{file=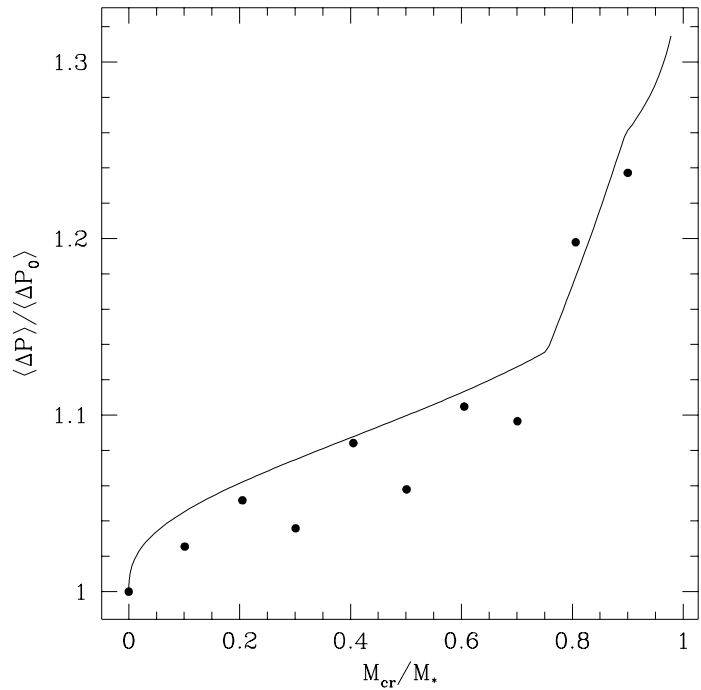,width=3.82in} \hfill}
\caption{The same as Figure \protect\ref{dp_x_schwz}b, except that periods
between 500 and 700 sec have been used to define the average period
spacing from the pulsation calculations; we have picked this range of
periods to mimic that observed in BPM 37093. For this case we see that mode
trapping effects result in larger deviations from the asymptotic relation.
\label{dp_x_led2}
}
\end{figure}

We now examine the more realistic case, where we include the modified
Ledoux criterion for the \bvfreq\, as described in \cite{Brassard91}
(1991). This plot is shown in Figure \ref{dp_x_schwz}b. Although the
overall shape of the plot has changed somewhat, the agreement between
the asymptotic and numerical results is still quite good. The observed
``kink'' for $0.75 \leq M\subs{cr}/\Mstar \leq 0.90$ is caused by
the oxygen mass-fraction decreasing from 0.80 to 0.00 in this range. If
we examine Figure \ref{pformg}a for the period formation region and we
imagine moving the crystallization region to the right, we see that as
we encounter the O/C transition zone, the rate of change of area under
the curve doubles, so we would expect the slope of the curve in Figure
\ref{dp_x_schwz}b to double as well, which is what we find.

If we use a smaller range of periods to define the period spacing
numerically, then we expect mode trapping effects to be amplified even
further. This is illustrated in Figure \ref{dp_x_led2}, where we have
used $\ell=2$ periods in the range 500--700 sec to calculate a period
spacing.  Thus, if we have a complete set of observed $\ell = 2$ periods
in this range, we can typically expect ``errors'' of order $\sim$5\%
in translating this to an asymptotic period spacing.

An equivalent statement to the period spacing increasing with
$M\subs{cr}/\Mstar$ is that the modes themselves are getting farther
apart in period, so their periods must also be increasing. To illustrate
this, we show how a spectrum of mode periods evolves continuously with
$M\subs{cr}/\Mstar$. Since mode identification between different models
is not a simple matter, we have calculated the spectrum of modes on a
fine enough mesh in $M\subs{cr}/\Mstar$ so that the period changes are
small compared to the differences between consecutive radial overtones.
We then identify a given mode at one mesh point with the nearest mode
in period of the neighboring mesh point.

\begin{figure}
{\hfill \epsfig{file=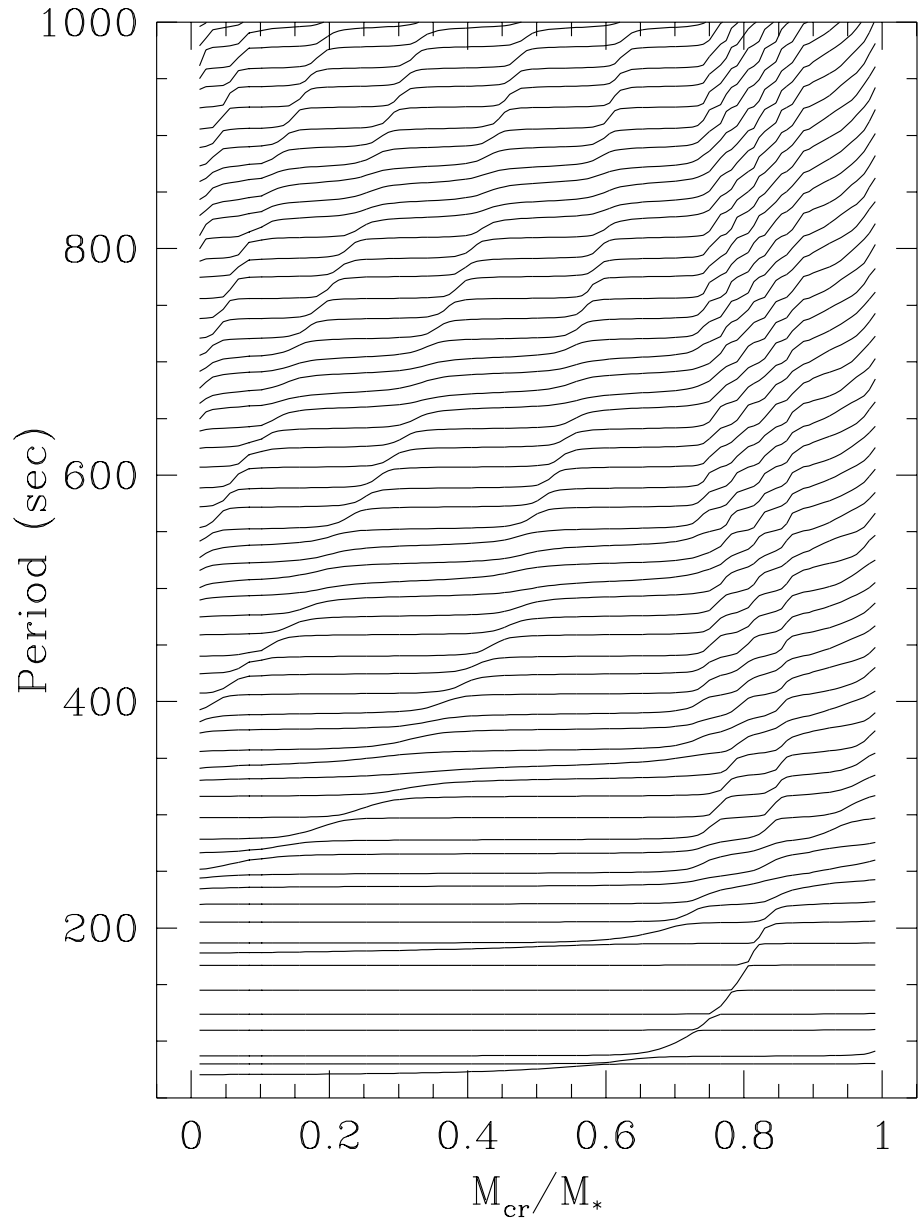,width=5.00in} \hfill}
\caption{The evolution of $\ell=2$ $g$-mode periods as a function of the
crystallized mass-fraction. We see that in a given region, the periods
are either increasing or relatively constant.
\label{avoid_l2_bw}
}
\end{figure}

The result of this calculation for $\ell=2$ periods is shown in
Figure \ref{avoid_l2_bw}, where the model considered is a 1.1
$\Msun$ with $T\subs{eff}=11,800$ K, $M\subs{He}/\Mstar=10^{-3}$, and
$M\subs{H}/\Mstar=10^{-5}$. We have used the ``hard-sphere'' approximation
for the solid/liquid boundary and the full Ledoux prescription for the
\bvfreq\ in calculating these periods. We have varied the parameter
$M\subs{cr}/\Mstar$ in increments of 0.01 from 0.00 to 0.99.  We note
that the periods either appear to be increasing or are relatively
constant. In fact, even in regions in which the period of a given mode
appears to be constant, its period is still slightly increasing with
$M\subs{cr}/\Mstar$.

Figure \ref{avoid_l2_bw} represents the most detailed calculation to date
showing how $g$-mode periods in white dwarf models evolve as a single
parameter is slowly varied (for a previous example, see \cite{Wood88}
1988). As such, it exhibits many interesting features. First, the ``kink''
in the periods in the range $0.75 \leq M\subs{cr}/\Mstar \leq 0.90$
is again due to the crystallized region moving out into a region with a
changing C/O profile; thus, this feature is merely a result of our assumed
C/O profile. A more intrinsic feature of this plot are the ``avoided
crossings.'' While it is never possible numerically to establish with
complete certainty whether or not true avoided crossings occur, the way
in which the periods evolve as a function of $M\subs{cr}/\Mstar$ is
strongly reminiscent of behavior found by \cite{Aizenman77} (1977). For
example, the two lowest period modes pictured in Figure \ref{avoid_l2_bw}
have what appears to be an avoided crossing at $M\subs{cr}/\Mstar
=0.58$. To the left of this point, the lower period mode has more of
its kinetic energy deep in the model near the solid/fluid interface,
while to the right of this point it is the higher period mode which has
its kinetic energy deeper.  Thus, the modes {\em do} switch character
at this point, in the manner found by \cite{Aizenman77} (1977).

Our general result that the $g$-mode periods increase due to the presence
of crystallization is not what was found by \cite{Hansen79} (1979), who
reported that the $g$-mode periods became shorter when the finite shear
of the solid core was included.  We believe that the resolution of this
disagreement lies in a re-interpretation of their calculated periods,
not in the periods themselves.  \cite{Hansen79} (1979) calculated the
periods of $k=1$ and 2 modes for $\ell=1$, 2, and 3, in both the fluid
case and in the case of a 99.9\% crystallized core. They found that in
the crystallized case, the $k=1$ periods had decreased by approximately a
factor of two compared to the fluid case; for example, the $\ell=1$ period
decreased from 193.8 sec to 99.8 sec. Our interpretation is that the
99.8 sec mode is actually a new mode, which would not exist if the core
were not crystallized. Thus, the main effect of the solid core in their
calculations was, in our view, to add an extra mode with a period below
that of the previous $k=1$ mode. To support this, we compare their $k=1$
periods in the fluid case with their $k=2$ periods in the solid case. For
$\ell=1$, 2, and 3, we find that their periods now {\em increase} from
193.8 to 193.9 sec, from 111.9 to 112.0 sec, and from 79.1 to 79.2 sec,
respectively. While these increases are small, they are consistent with
what one might expect from a $T\subs{eff} \sim$ 10,000~K Fe core white
dwarf model which is strongly degenerate in its interior. In addition,
the periods in the uncrystallized and the crystallized state are close
enough to strengthen our conviction that this is actually the ``correct''
mode identification.

Using our ``global'' code, we are numerically unable to treat models
which are more than 97\% crystallized. For 97\% crystallized models, we
do find evidence for low-period ``interfacial'' modes which do not exist
in the uncrystallized case; interfacial modes such as these were found
in neutron star models by \cite{McDermott88} (1988). These modes could
be the new modes found by \cite{Hansen79} (1979). We caution, however,
that we do not understand the properties of these modes, i.e., how they
change period as the degree of crystallization changes and whether or not
the standard definition of radial overtone number is still meaningful. We
are therefore unable to extrapolate these results with confidence to
the case of 99.9\% crystallization which \cite{Hansen79} treated.

\section{$\langle \Delta P \rangle$ as a Function of the Model Parameters}

In uncrystallized models, the period spacing is a function of many
things, including the total stellar mass, the effective temperature,
and the hydrogen layer mass. This is still true in the crystallized
case, and we examine the effects which each has on $\langle \Delta P
\rangle$.  The fiducial model against which we compare our calculations
is a model with $\Mstar = 1.1 \Msun$, $T\subs{eff} = 11,800$ K,
$M\subs{H}/M_{\star}=10^{-5}$, and $M\subs{He}/M_{\star}=10^{-3}$.
Unless otherwise stated, all periods are calculated using the modified
Ledoux prescription for the \bvfreq.

\subsection{The Hydrogen Layer Mass, $M\subs{H}$}

For 0.6 $M_{\odot}$ models, nuclear burning considerations force
$M\subs{H}/M_{\star}$ to be smaller than a few times $10^{-4}$
(\cite{Iben84} 1984; \cite{Iben85} 1985).  For models near  1.1
$M_{\odot}$, this translates into $M\subs{H}/M_{\star} \lesssim 10^{-5}$
due to the higher gravities and pressures.  We therefore examine models
with  $M\subs{H}/M_{\star}$ between $10^{-10}$ and $10^{-5}$.

\begin{figure}
\plottwo{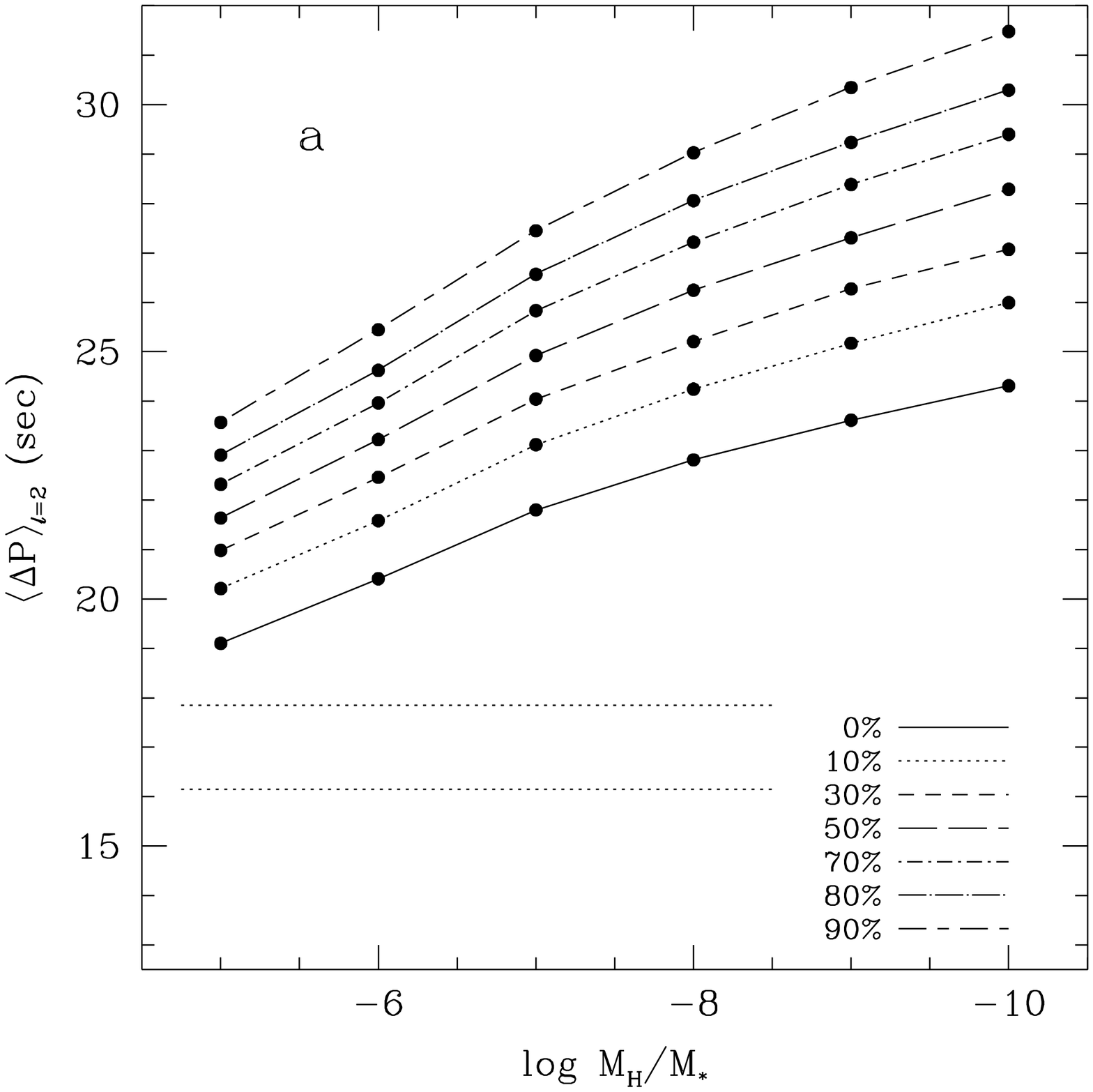}{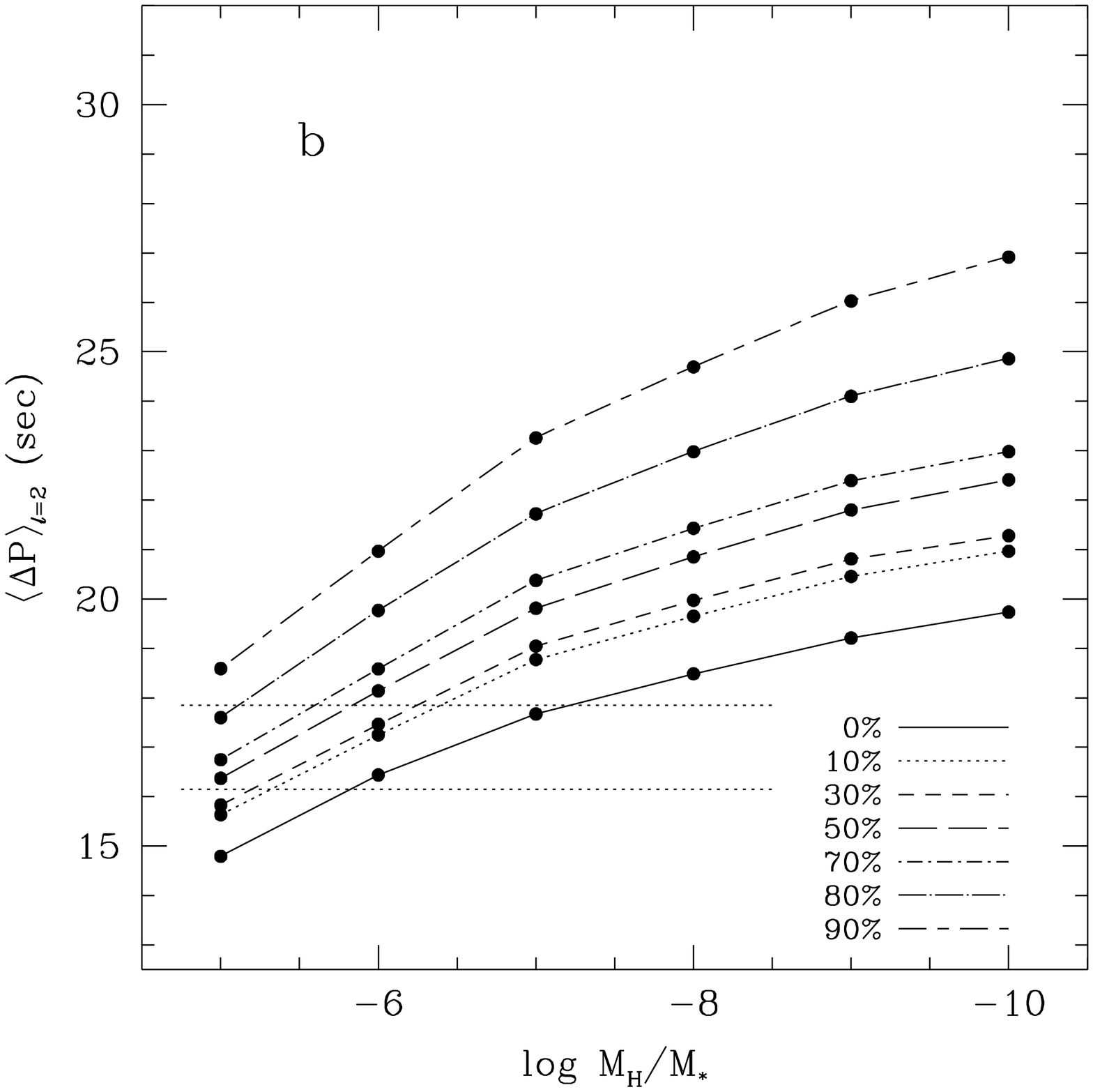}
\caption{(a) $\langle \Delta P \rangle$ as a function of $\log
M\subs{H}/M_{\star}$ for differing degrees of crystallization with $N^2$
calculated according to the Schwarzschild criterion. (b) The same as
(a), but with $N^2$ calculated using the Ledoux prescription.
\label{dpschwz}
}
\end{figure}

In Figure \ref{dpschwz}a, we plot $\langle \Delta P \rangle$ versus
$\log M\subs{H}/M_{\star}$ for different degrees of crystallization,
as shown in the legend. For this model, we have used a C/O core and set
$M\subs{He}/M_{\star} = 10^{-3}$ and $T\subs{eff}=11,800$~K, and we have
calculated the \bvfreq\ using the Schwarzschild criterion, so that we
may minimize mode trapping effects as much as possible. We see that the
effect of increasing crystallization is to increase $\langle \Delta P
\rangle$ at all compositions. Similarly, the effect of decreasing $\log
M\subs{H}/M_{\star}$ is also to increase $\langle \Delta P \rangle$,
for all degrees of crystallization. Thus, a change in one can mimic a
change in the other.  Figure \ref{dpschwz}b shows the more physical case
where we have used the Ledoux prescription for calculating the \bvfreq\
in this model. The same trends are still evident, but the period spacing
itself has decreased by 3--4 sec for all the models. This difference is
due to the nontrivial contribution of the composition transition zones.

The horizontal dashed lines in Figures \ref{dpschwz}a and \ref{dpschwz}b
are useful in demonstrating how observations could be used to constrain
the parameter space. Here, we assume a ``measured'' value of $\langle
\Delta P \rangle \sim 17$~sec. The dashed lines represent an uncertainty
of 5\% in translating this ``observed'' $\langle \Delta P \rangle$ to
an asymptotic value, as is suggested by the deviations due to a finite
sampling of the period range in Figure \ref{dp_x_led2}.  For these
calculations, we have calculated the period spacing between consecutive
$\ell=2$ modes, all with $m = 0$. The exact same dependencies hold for
the case of $\ell=1$ modes, if the mean period spacings are multiplied
by a factor of $\sqrt{3}$.

From Figure \ref{dpschwz}b, we find the following constraints on our
parameter space: $-7 \leq \log M\subs{H}/M_{\star} \leq -5$ and $ 0.00
\leq M\subs{cr}/M_{\star} \leq 0.80$. This is a fairly large range
for each parameter, but they are now no longer independent. If we know
one of them, then that can reduce the allowed range for the other. For
instance, if the model is 50\% crystallized, then we must have $-6 <
\log M\subs{H}/M_{\star} < -5$.  Additionally, from Figure \ref{dpschwz}a
we see that there is no choice of parameters for which the period spacing
matches the ``observed'' value. This demonstrates the large effect which
the composition transition zones have on the average period spacing.

\subsection{The Total Stellar Mass, $\Mstar$}

\begin{figure}
\plottwo{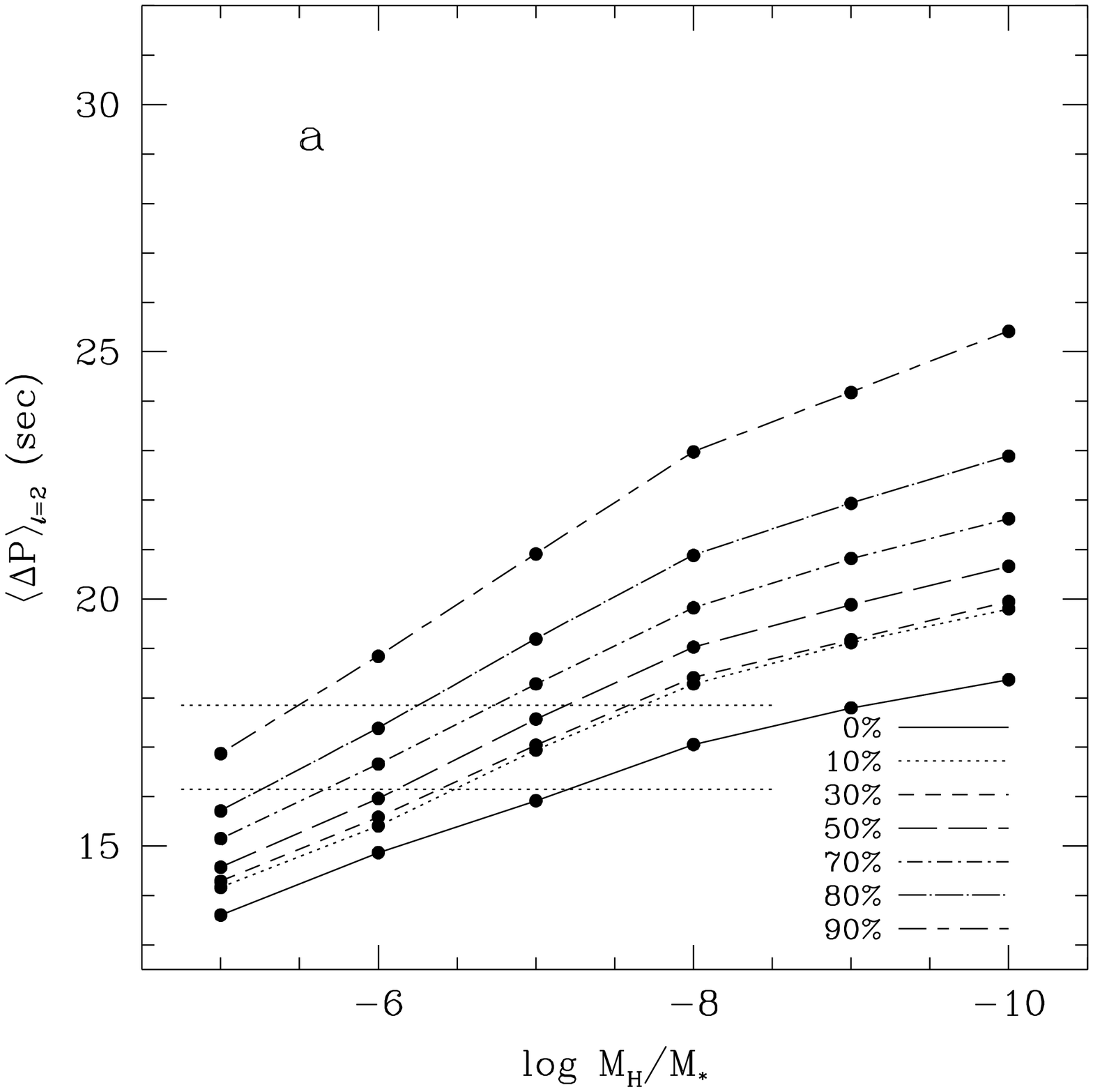}{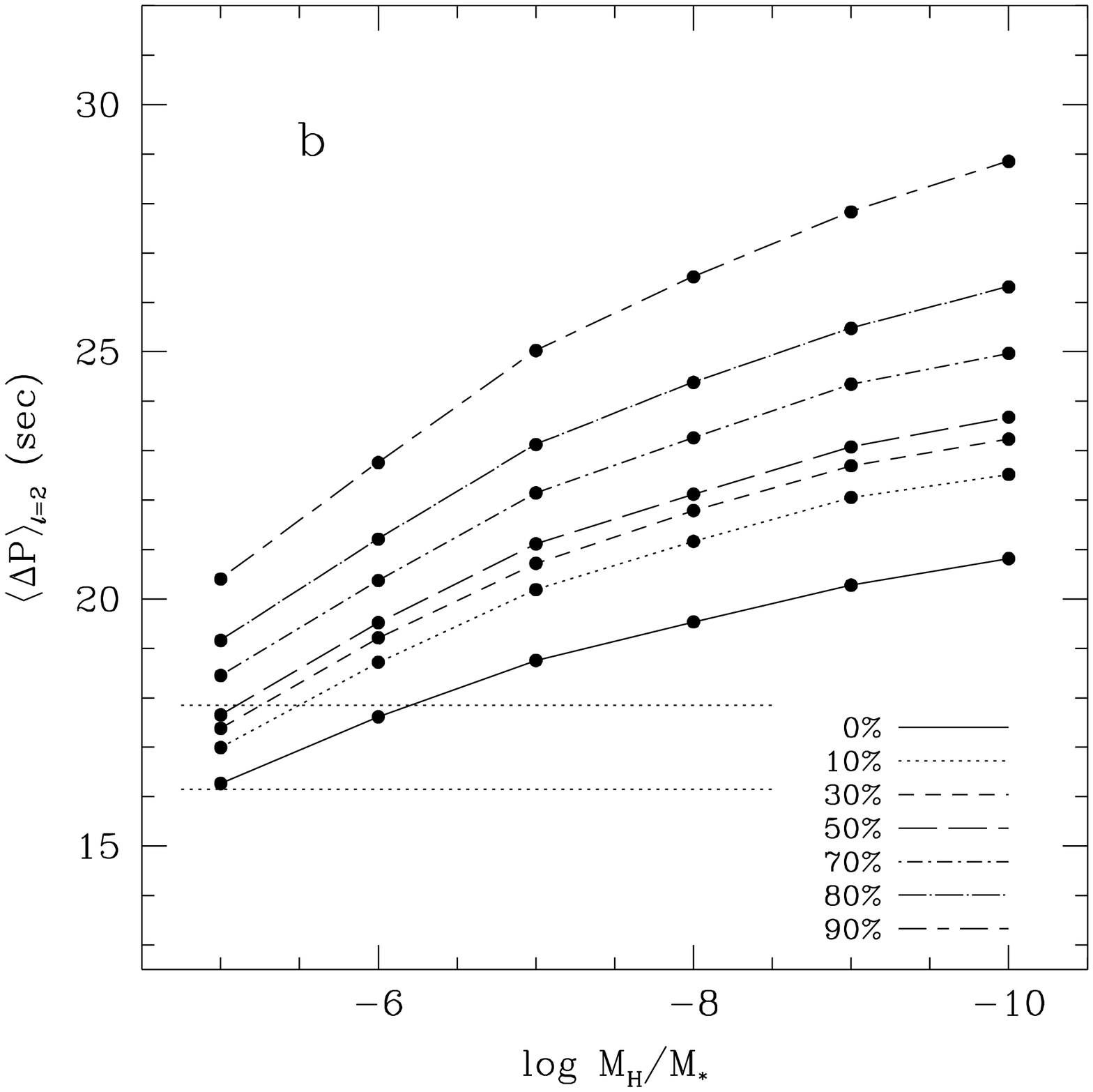}
\caption{The same as Figure \protect\ref{dpschwz}b, but for (a)
1.15~$\Msun$ and (b) 1.05~$\Msun$ models.
\label{dpfull1.15}
}
\end{figure}

We now consider models which differ only in mass from our fiducial model;
all the other parameters are held fixed.  In Figure \ref{dpfull1.15}a
we plot the average period spacing for a set of $\Mstar =$ 1.15 $\Msun$
models, again as a function of $M\subs{H}$, where we continue to use the
more physical Ledoux prescription for the the \bvfreq. Since the more
massive models are smaller in radius, they have a higher average density,
and therefore smaller periods and period spacings. For the less massive,
1.05 $\Msun$ models in Figure \ref{dpfull1.15}b, we find the opposite is
the case;  these models are larger in radius and therefore have larger
period spacings.

\subsection{The Effective Temperature, $T\subs{eff}$}

\begin{figure}
{\hfill \epsfig{file=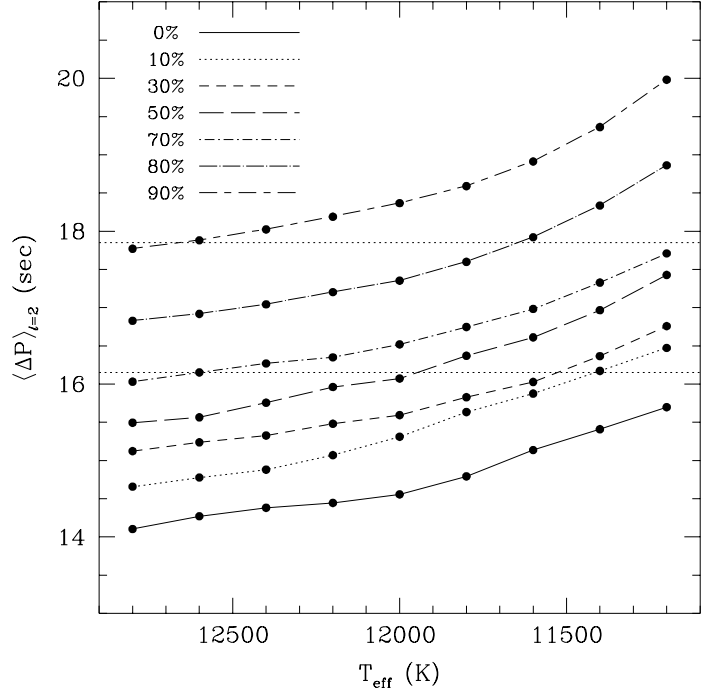,width=3.82in} \hfill}
\caption{The average period spacing as a function of $T\subs{eff}$
for different degrees of crystallization, as shown in the legend. The
models all have $\Mstar = 1.1 \, \Msun$, $M\subs{He}/\Mstar = 10^{-3}$,
and $M\subs{H}/\Mstar = 10^{-5}$.
\label{dpteff1.1}
}
\end{figure}

In Figure \ref{dpteff1.1} we show how the mean period spacing for $\ell=2$
modes varies as a function of the effective temperature of our fiducial
models. The horizontal dotted lines again bracket an ``observed''
period spacing of 17 sec.  We see that as the models cool, the period
spacing increases.  This occurs because the models are becoming more
degenerate. As the models approach complete degeneracy, the \bvfreq\
becomes arbitrarily small, except in composition transition zones,
so the periods and period spacings become large.

\subsection{Scaling Relations}

The results of the previous sections may be used to obtain approximate
scaling relations for $\langle \Delta P \rangle$. Using models with
$T\subs{eff}$ between 11,200 K and 12,800 K, $\Mstar$ between 1.05
$\Msun$ and 1.15 $\Msun$, and hydrogen layer masses with $H \equiv -\log
M\subs{H}/\Mstar$ between 5 and 10 (all ranges inclusive), we obtain
the following relation:
\begin{equation}
\langle \Delta P \rangle_{\ell=2} = A \bar{f} [1+0.54 \bar{f} (H-5)]^{0.24}
\bar{M}_{\star}^{-1.7} \bar{T}\subs{eff}^{-0.95},
\label{dpfit}
\end{equation}
where $A \equiv 14.7$ sec, $\bar{M}_{\star} \equiv M_{\star}/(1.1 \Msun),
\bar{T}\subs{eff} \equiv T\subs{eff}/(12,000$\,K), and $\bar{f} \equiv
\langle \Delta P \rangle/\langle \Delta P \rangle_0$ is the ratio of
the asymptotic period spacing at finite crystallization to that at zero
crystallization, e.g., the solid line in Figure~\ref{dp_x_schwz}b. The
bar on $f$ indicates that we have chosen $f$ for an ``average''
model, where by average we mean a model which has $M_{\star}=1.1
\Msun, T\subs{eff}=12,000$ K, $M\subs{H}/\Mstar = 10^{-5}$, and
$M\subs{He}/\Mstar = 10^{-3}$.

Next, we examine the sensitivity of $\langle \Delta P \rangle$ to small
changes in these parameters. If we look at small variations around a model
which is 50\% crystallized and has a $H$ layer thickness corresponding
to $H = 5$, we find
\begin{equation}
\frac{\delta \langle \Delta P \rangle}{\langle \Delta P \rangle}
= 0.13\, \delta m\subs{cr} + 0.15\, \delta H 
  - 1.70 \frac{\delta \Mstar}{\Mstar}
  -0.95 \frac{\delta T\subs{eff}}{T\subs{eff}},
\end{equation}
where we have defined $m\subs{cr} \equiv M\subs{cr}/\Mstar$, and $\delta
Y$ represents a small change in a given quantity $Y$.  From fits of
spectra of BPM 37093, \cite{Bergeron95} (1995) find $T\subs{eff}=11,740
\pm 200$ K, and $\Mstar = 1.09 \pm 0.05 \Msun$. From this we see that
the errors in the mass determination produce about 5 times the effect
of the errors in the temperature determination. Thus, $\Mstar$ is the
most important input parameter which the observations can provide. The
quantities $m\subs{cr}$ and $H$, the crystallized mass fraction and
the negative of the log of the hydrogen layer mass, respectively,
are not observable quantities in the standard sense. They can only be
determined from an asteroseismological analysis of a particular star,
which leads us to the topic of the next section.

\section{Mode Trapping}

The traditional way to obtain information about the surface layer
thicknesses of white dwarfs is to use mode trapping information for
individual modes, i.e., calculate $\Delta P_k \equiv P_{k+1} - P_k$
directly from the data set and match this to numerical calculations. There
is no reason why this will not work now, as long as we have enough
well-identified consecutive overtones.

In general, a transition zone may trap a mode in the region above it
{\em or} below it.  For a mode to be trapped in the outer hydrogen
layer, it needs to have a resonance with the He/H transition region
such that its vertical and horizontal displacements both have a node
near this interface (\cite{Brassard92} 1992); this is the case which is
traditionally referred to as mode trapping in the context of white dwarfs.
If we imagine integrating this mode inward from the surface using the
boundary conditions there, then we see that all this condition depends
on is the mode frequency.  Whether or not a frequency which would be
trapped is indeed an allowable normal mode frequency {\em does} depend
on the amount of crystallization in the core.  From this, we see that
it should be possible to disentangle the effects of crystallization and
mode trapping.

More precisely, \cite{Brassard92} (1992) find that the average period
difference $\langle \Delta P \rangle_t$ between successively trapped modes is
\begin{equation}
\langle \Delta P \rangle_t = \frac{2 \pi^2}{\sqrt {\ell (\ell+1)} }
   \left[\int_{rH}^{r2} N dr/r \right]^{-1},
\label{asymph}
\end{equation}
where $rH$ is defined as the radius at the base of the hydrogen layer; the
integral is therefore over the hydrogen surface layer only. We see that
this does not depend on any of the properties of crystallized region,
but only on those of the hydrogen envelope. 

\begin{figure}
\plotone{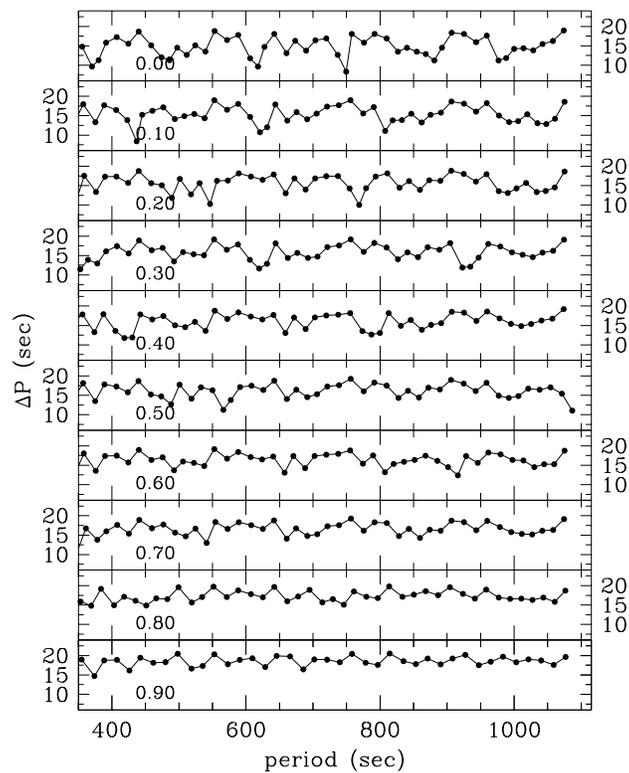}
\caption{$\Delta P_i \equiv P_{i+1} - P_i$ (forward period difference)
versus $P_i$ (period) for $\ell=2$ modes. Each panel is labelled by the
degree of crystallization assumed for the model, with the other model
parameters being held constant.
\label{pdif_l2}
}
\end{figure}

In Figure \ref{pdif_l2}, we plot the forward period difference, $\Delta
P_i \equiv P_{i+1} - P_i$, versus period, $P_i$, for an equilibrium
model with $M_{\star}=1.1 \Msun, T\subs{eff}=11,800$ K, $M\subs{H}/\Mstar
= 10^{-5}$, and $M\subs{He}/\Mstar = 10^{-3}$. This shows how the mode
trapping changes as the crystallized mass fraction is varied from 0.0
to 0.9 in increments of 0.1.  In general, we see that the amplitude
(strength) of the trapping decreases with increasing crystallization. This
is because as the degree of crystallization increases, all modes become
more like envelope modes, which decreases the differences between the
trapped and untrapped modes.

Concerning the detailed structure of the mode trapping itself, the
combined effect of the different transition zones makes it difficult
to define a trapping cycle.  Furthermore, we see that this structure
changes significantly as the degree of crystallization is changed by
only 10\%. This suggests that we will need to examine the degree of
crystallization in smaller increments.

\begin{figure}
\plottwo{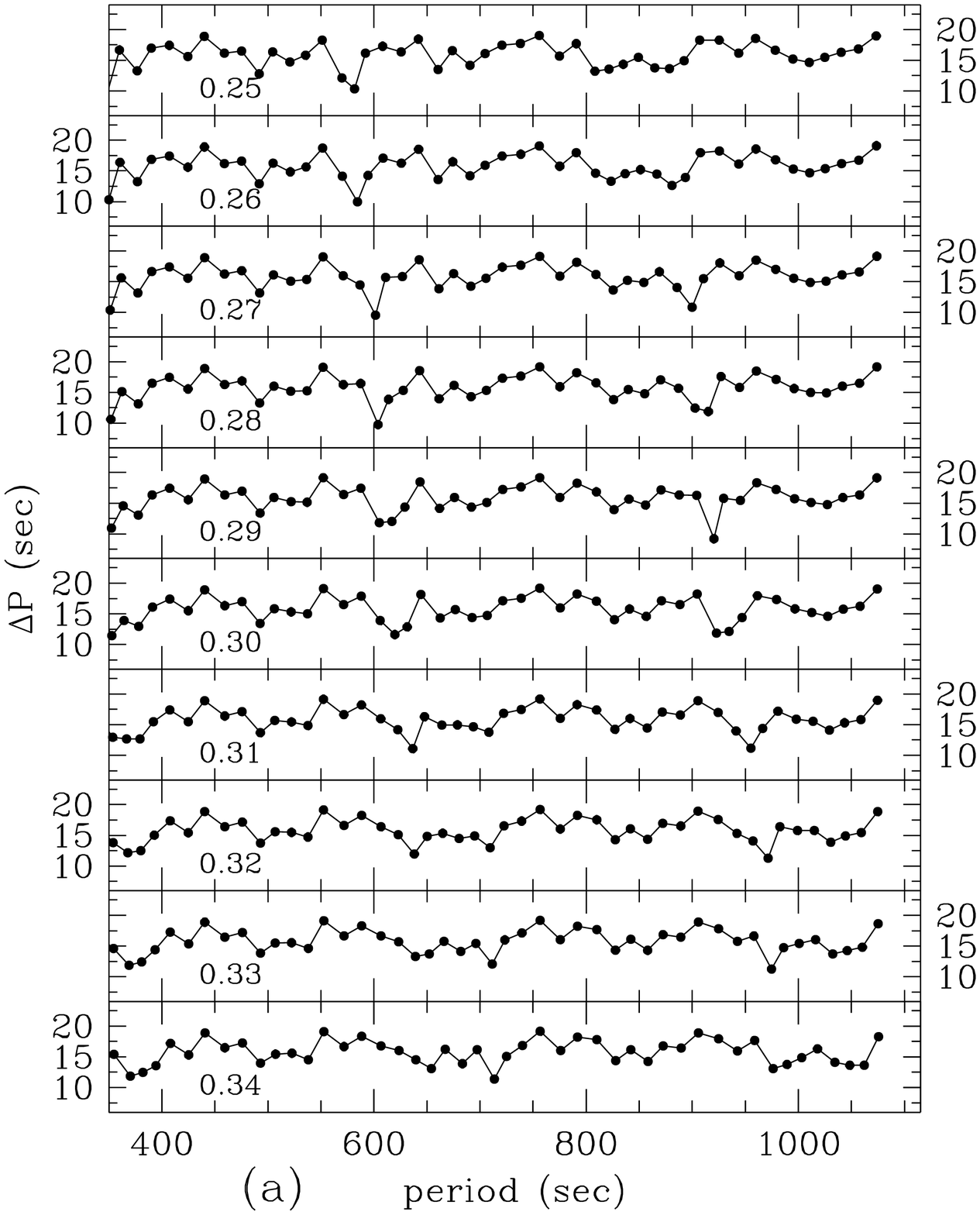}{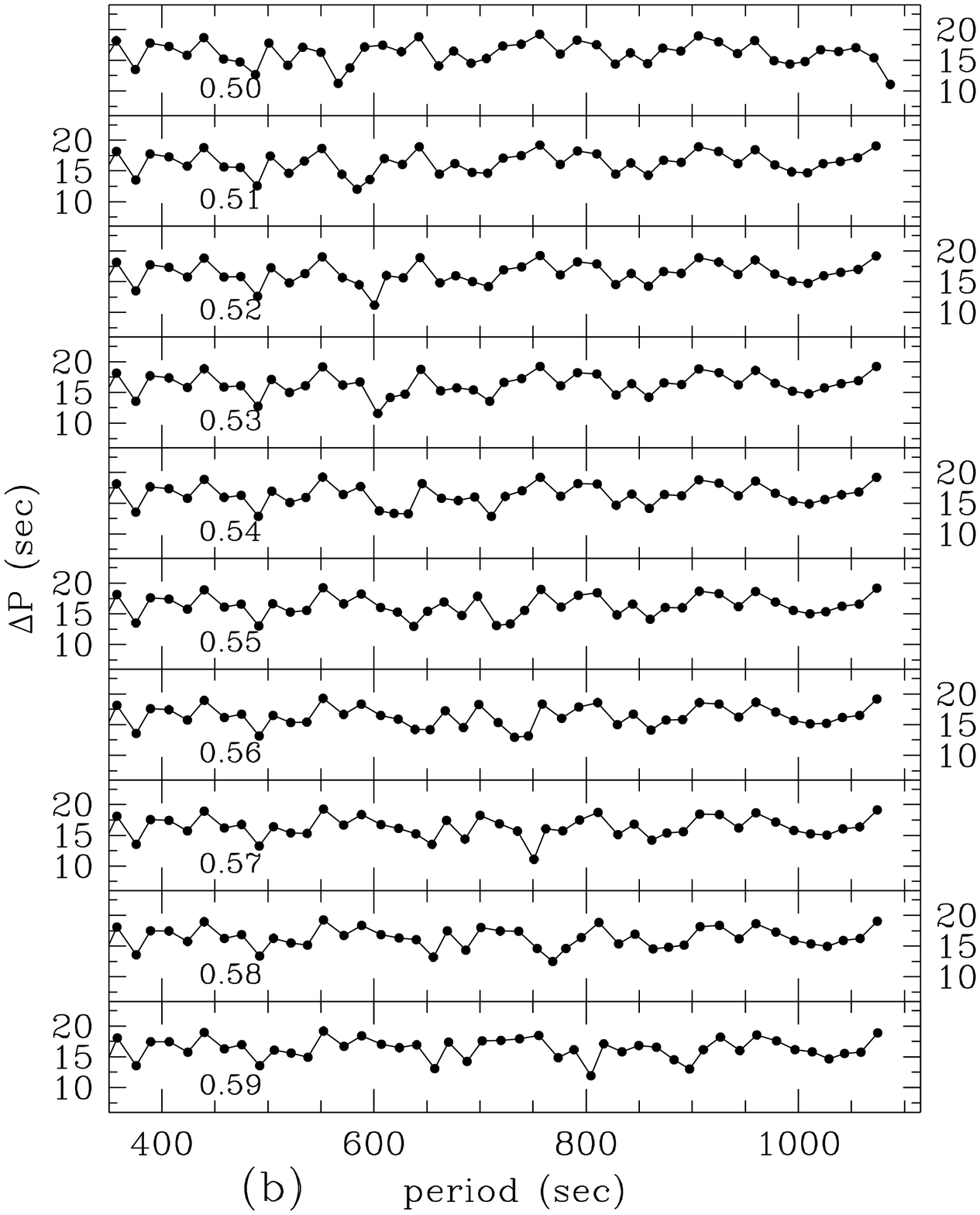}
\caption{The filled circles connected by lines show the period spacing
in the model versus the period, for degrees of crystallization varying
between (a) 25\% and 34\% and between (b) 50\% and 59\%. 
\label{pdif1_l2}
}
\end{figure}

Figure \ref{pdif1_l2}a is a more detailed version of Figure \ref{pdif_l2}
which shows how $\Delta P$ changes as the crystallized mass fraction
is increased from 0.25 to 0.34 in increments of 0.01. First, we note
that there are many trapping features which move uniformly to the
right as the the degree of crystallization is increased. For instance,
there is a trapping feature with a period of $\sim 580$~sec at 25\%
crystallization which migrates to a period range of $\sim 640$~sec at
32\% crystallization.  Second, there are many features which remain
relatively constant. The mode trapping structure in the range 420--500
sec is virtually unchanged, and the mode with a period of $\sim 775$
sec is also somewhat trapped in the majority of the panels.  This $\sim$
775 sec mode has a period which does not evolve as rapidly as many of the
other modes. Even so, its period changes by $\sim$ 0.6 sec for every 5\%
change in the degree of crystallization.  Since it is possible to measure
periods to quite high accuracies of a few tenths of a second (e.g.,
\cite{Winget91} 1991, \nocite{Winget94} 1994), we should in principle,
using modes such as this as well as more sensitive modes, be able to
derive quite accurate estimates of the crystallized mass fraction,
if we are able to obtain a unique solution.

Figure \ref{pdif1_l2}b shows a different range of crystallization, this
time between 50\% and 59\%, for the same white dwarf model as was used
for Figures \ref{pdif_l2} and \ref{pdif1_l2}a. Here, we see that there
is again a trapping feature which migrates through the 600 sec region,
as well as a mode with a period again at $\sim$ 775 sec which tends to
be trapped (at least in the upper 7 panels). The fact that some periods
are strongly affected by changes in the crystallized mass fraction while
others are not suggests that we may be able to disentangle the effects
of different surface layer masses from those due to crystallization.

\section{Objective Fitting Procedures}
We need an automated procedure for searching parameter space, both
to obtain more precise fits and to address the issue of uniqueness
of fit. The sensitivity of the trapping features to the crystallized
mass fraction is both a blessing and a bane: It is a blessing because
this should allow us in principle to determine precise values for
$M\subs{cr}/\Mstar$, and it is a bane because in practice it requires
the computation of an enormous number of models on a fine grid in order
to sample the parameter space adequately.

We are currently exploring different methods which would address these
issues. The first part of the problem is choosing a method such as
``simulated annealing'' or a ``genetic algorithm'' which can find global
minima of multidimensional functions. The second part is automatically
generating the equilibrium models with a given set of fit parameters,
so that these models can be examined pulsationally. Traditionally, the
evolution of such models has been a ``hands on'' procedure, and this is
true of our evolutionary models as well.


Unfortunately, this problem is beyond the scope of this paper. Given the
recent increase in both processor speeds and the degree to which problems
are being parallelized, it should be possible to implement an objective
fitting scheme which will allow us to sample adequately the parameter
space of the models. Such an approach is currently being developed at
the University of Texas.  This should, among other things, allow us to
assess objectively the uniqueness of our asteroseismological fits.

\section{Conclusions}

In this paper we have calculated the effect which crystallization in
the cores of our white dwarf models has on the frequency spectrum of
pulsations. To a very high degree of accuracy, we find that the kinetic
energy of the $g$-modes is excluded from the crystallized cores of our
models. As the degree of crystallization is increased, the kinetic energy
of these modes is pushed farther out from the central regions, and both
the periods and the mean period spacing $\langle \Delta P \rangle$
between consecutive radial overtones of the same $\ell$ increases.
Using an ``observed'' value of $\langle \Delta P \rangle_{\ell=2}$ =
17 sec, we show how the range of possible models can be constrained, and
how mode trapping features may be used to obtain more precise information
about these fits. Since some periods are strongly affected by changes
in the crystallized mass fraction while others are not, we may be able
to disentangle the effects of different surface layer masses from those
due to crystallization.

The introduction of a crystalline medium which is able to support shear
does allow a new class of modes to exist, the torsional or $t$-modes.
Since these modes have zero radial displacement, they should be unable
to couple to the overlying fluid layers and should therefore remain
unobservable. The $p$-modes have periods which are only a few percent
different from their uncrystallized values. Since these modes are not
observed to be excited, they are also not of interest in the context of
BPM 37093 and the other ZZ Ceti's.

By investigating stars such as BPM 37093, asteroseismology may eventually
be able to tell us whether crystallization occurs in the way we expect
theoretically. Since crystallization and the effects of phase separation
are the largest single sources of systematic uncertainties in the white
dwarf luminosity function, this would allow us to improve our estimates
of the age of the Galactic disk as derived from the observed white dwarf
luminosity function. Furthermore, since phase separation, if it occurs,
affects the composition of the central regions of white dwarfs, this could
systematically affect the observed characteristics of SNIa (e.g., total
luminosity), which are believed to come from white dwarf progenitors.

Finally, digital surveys now in progress promise to add considerably to
the presently known number of cool white dwarfs. For instance, the Sloan
Digital Sky Survey (\cite{Gunn95} 1995) should increase this number by
a factor of approximately 20, with the result that we may have 20 such
stars with which to test the theory of crystallization.

\section{Acknowledgments}

We would like to thank C. J. Hansen for many insights and valuable
discussions, as well as the referee for his helpful comments.

This work was supported in part by the National Science Foundation
under grant AST-9315461 and by the NASA Astrophysics Theory Program
under grant NAG5-2818.

\appendix

\section{Spheroidal Oscillation Equations in a Crystalline Medium}
\label{derivation}
\subsection{The Equations}

\setcounter{footnote}{0}

We define $\xi_r$ and $\xi_h$, the radial and horizontal parts of the total
displacement, in terms of the total vector displacement, i.e.,
\begin{equation}
\vec{\xi} = \left[\xi_{r}(r),\xi_h(r)\frac{\partial}{\partial \theta},
\xi_h(r)\frac{1}{\sin\theta}\frac{\partial}{\partial \phi}\right] \ylm.
\label{eigensep}
\end{equation}
We take the following equations from \cite{Hansen79} (1979).
The oscillation variables are
\[ z_1=\frac{\xi_r}{r}, \]
\[ z_2= \frac{1}{\mu_0} \left(\lambda \alpha + 
    2 \mu \frac{d \xi_r}{dr}\right), \]
\[ z_3 = \frac{\xi_h}{r}, \]
and
\[ z_4 = \frac{\mu}{\mu_0} \left(\frac{d \xi_h}{dr}-\frac{\xi_h}{r}+
  \frac{\xi_r}{r}\right), \]
where $\xi_r$ and $\xi_h$ are the radial and horizontal displacements,
respectively, as defined in equation \ref{eigensep}, $r$ is the radius,
$\lambda = \Gamma_1 p - \frac{2}{3} \mu$, $\mu$ is the shear modulus,
$\alpha \equiv \frac{1}{r^2} \frac{d}{dr}\left(r^2 \xi_r\right)
 - \lhat \frac{\xi_h}{r}$, with $\lhat \equiv \ell (\ell+1)$.
These variables are the same as those in \cite{Hansen79} (1979) except
that we have divided $z_2$ and $z_4$ by $\mu_0 \equiv \mu(r=0)$, so that
the equations are dimensionless.\footnote{We note that there is a
typographical error in the definition of $z_4$ in \cite{Hansen79}
(1979)---an additional factor of $1/r$---but it does not propagate
throughout the rest of their formulae.} The fourth order system of
equations (in the Cowling approximation) is then 
\begin{eqnarray*}
  r z_1^{\prime} & = & -(1+2 \lambda \delta) z_1 + \mu_0 \delta z_2 
  + \lambda \lhat z_3, \\
  r z_2^{\prime} & = & \frac{1}{\mu_0}(-\sigma^2 \rho r^2 - 4 \rho g r + 
    4 \pi G \rho^2 r^2 + 4 \mu \beta \delta) z_1  
    - 4 \mu \delta z_2 
    + \frac{\lhat}{\mu_0} (\rho g r-2 \mu \beta \delta) z_3 + \lhat z_4, \\
  r z_3^{\prime} & = & -z_1 + \frac{\mu_0}{\mu} z_4, \\
  r z_4^{\prime} & = & \frac{1}{\mu_0} (g \rho r-2 \mu \beta \delta) z_1 - 
    \lambda \delta z_2 
    + \frac{1}{\mu_0} \left\{-\rho \sigma^2 r^2 +
    2 \mu \delta \left[\lambda (2 \lhat-1)+2 \mu (\lhat-1)\right]\right\} z_3
    -3 z_4,
\end{eqnarray*}
where the prime denotes $\frac{d}{dr}$, $\delta \equiv (\lambda+2
\mu)^{-1}$, $\beta \equiv 3 \lambda+2 \mu$, $g$ is the acceleration due to
gravity, and $\rho$ is the density.

\subsection{Central Boundary Conditions}
Since the models we are considering are crystallized in the center, we
need to obtain the boundary conditions in the center so that we may begin
the outward integrations. If we assume that the solutions go like
$r^s$ near the center, we find four solutions:
$s = \ell-2,\ell,-(\ell+1),-(\ell+3)$. Only the first two solutions are
regular at the origin, so they span the space of physical solutions. 
The general solution near the center is therefore given by
\[
\{z_i\} =
a
\left(\begin{array}{c}
    1 \\[.5em]
    2 (\ell-1) \\[.5em]
    1/\ell \\[.5em]
    2 (\ell-1)/\ell
\end{array}\right)
r^{\ell-2} +
b
\left(\begin{array}{c}
    \frac{(\ell+1)[\lambda \ell+\mu (\ell-2)]}{
      2[\lambda \ell(\ell+2)+\mu(\ell^2+2 \ell-1)]} \\[.5em]
    \frac{(\ell+1)[\lambda (\ell^2-\ell-3)+\mu(\ell^2-\ell-2)]}{
      \lambda \ell(\ell+2)+\mu(\ell^2+2 \ell-1)}\\[.5em]
    \frac{\lambda (\ell+3)+\mu(\ell+5)}{
      2[\lambda \ell(\ell+2)+\mu(\ell^2+2 \ell-1)]}\\[.5em]
    1
\end{array}\right)
r^{\ell},
\]
where $a$ and $b$ are arbitrary coefficients and where $\mu$ and $\lambda$
in the above formula are taken to have their central values. These two
solutions for the eigenfunction near the center are equivalent to the
relations given in \cite{Crossley75} (1975), if the Cowling approximation
is used.

\subsection{The Solid/Fluid Interface}

In practice, we integrate each independent solution outward from the
center. With the exception of $z_3$, the $\{z_i\}$ are continuous at 
the solid/fluid interface. Since $z_4 = 0$ in the fluid, we choose the ratio of 
$a$ and $b$ such that $z_4$ vanishes at this interface. This leaves only
one overall normalization constant. Furthermore, $y_1 = z_1$ at the boundary.
Since $z_2$ is also continuous, we have
\[z_2 = \lambda \alpha/\mu_0 = \lambda V_g (y_1-y_2)/\mu_0, \]
where we have used the oscillation equations in the fluid to express
$\alpha$ in terms of the Dziembowski variables $\{y_i\}$.
At the fluid/solid interface, if we solve for the $\{y_i\}$
in the fluid in terms of the $\{z_i\}$ in the solid then we find
\begin{eqnarray*}
  y_1 & = & z_1, \\
  y_2 & = & z_1 - \frac{\mu_0}{\lambda V_g} z_2, 
\end{eqnarray*}
where $V_g = g r/c_s^2$, and $\lambda$ is now $\Gamma_1 P$ since $\mu$
is zero in the fluid. Since we now have specified $y_1$ and $y_2$ (up to
an overall normalization constant which is present in the $\{z_i\}$),
we can now integrate the normal oscillation equations in the fluid
(in the Cowling approximation) out to the photosphere of the model.

The main difficulty in applying this procedure is that numerical noise
can come to dominate the integrations in the crystalline core.  The model
which \cite{Hansen79} (1979) considered was a pure Fe core model near
10,000 K. As a result, the theory of crystallization suggested it should
be about 99.9\% crystallized by mass. The technique which we have used
would probably not be viable for this case. The problem is that the two
independent solutions, while quite different near the center, become
almost linearly dependent farther out. Thus, we lose the ability to
calculate the ``difference'' between the two solutions which is needed
in order to set $z_4$ equal to zero at the solid/fluid interface. For
our program, numerical noise dominates this process for $g$-modes in
models which are more than 98\% crystallized.

In terms of the physics, however, we are somewhat over-dramatizing the
situation, since nearly all of the pulsational results in this paper
are based on the simple approximation that $y_1=0$ at the solid/fluid
boundary. From the self-consistent treatment, we have found this to
be an extremely good approximation from 0\% crystallization to 98\%
crystallization, and we have no reason to believe this situation will
change at higher amounts of crystallization. Using this simplified
treatment ($y_1=0$ at the solid/fluid boundary), we are therefore able
to treat accurately arbitrary degrees of crystallization.

\section{The Boundary Conditions on $\Phi'(r)$ at the Crystallization Boundary}
\label{boundary}

The perturbations to the gravitational potential are, of course, generated by
the perturbations in the density. They may therefore be written as
\begin{equation}
 \Phi'({\bf r}) = G\int dV' \frac{\rho'({\bf r}')}{
  \left|{\bf r-r' }\right|},
\end{equation}
where we have assumed that the density vanishes at the surface. If we now
write $\Phi'({\bf r})=\Phi'(r) \ylm$ and $\rho'({\bf r}')=\rho'(r') 
Y_\ell^m(\theta',\phi')$ and use the usual expansion of $\left|{\bf r-r' }\right|$ in surface harmonics, then we arrive at the result of 
\cite{Christensen-Dalsgaard76} (1976):
\begin{equation}
\Phi'(r) = \frac{4\pi G}{2 \ell+1} 
    \left\{ r^{-(\ell+1)} \int_0^r \rho'(r') r'^{\ell+2} dr' 
    + r^{\ell} \int_r^{R} \rho'(r') r'^{-\ell+1} dr'
    \right\}.
\label{pert}
\end{equation}

We will assume that there is no motion for $r<r\subs{x}$, where
$r\subs{x}$ is the radius of the crystallization boundary. Thus, we have
$\rho'(r)=0$ for $r<r\subs{x}$. If we now take a derivative of $\Phi'(r)$
in the region $r\subs{x} < r <R$, we find
\[
\frac{d\Phi'(r)}{dr} = \frac{4\pi G}{2 \ell+1} 
    \left\{-(\ell+1) \, r^{-(\ell+2)} \int_0^r \rho'(r') r'^{\ell+2} dr' 
    + \ell \, r^{\ell-1} \int_r^{R} \rho'(r') r'^{-\ell+1} dr'
    + r \rho'(r) - r \rho'(r)
    \right\},
\]
where we have assumed that the density is zero at the outer boundary.
Evaluating this at $r=r\subs{x}$, and remembering that $\rho'(r)=0$
for $r<r\subs{x}$, we find that the first integral vanishes, which,
along with the cancellation of the last two terms, leaves only
\begin{equation}
\frac{d\Phi'(r\subs{x})}{dr} = \frac{4\pi G}{2 \ell+1} 
     \, \ell \, r_{r\subs{x}}^{\ell-1} \int_{r\subs{x}}^{R} 
      \rho'(r') r'^{-\ell+1} dr'.
\label{pert2}
\end{equation}
Finally, using equation \ref{pert} to evaluate $\Phi'(r)$ at $r = r\subs{x}$, and 
combining it with equation \ref{pert2}, we obtain the final result:
\begin{equation}
\frac{d\Phi'(r\subs{x})}{dr} = \frac{\ell}{r\subs{x}} \, \Phi'(r\subs{x}).
\end{equation}
This is the same boundary condition as is usually encountered in the
uncrystallized case for $r$ approaching zero (e.g., see \cite{Unno89} 1989); 
we see that it is unchanged by the presence of a rigid, crystallized core.

\section{Global Conservation of Momentum for $\ell=1$ Modes}

As a note, we mention that the derivation by
\cite{Christensen-Dalsgaard76} (1976) that $\ell$ = 1 modes conserve
the total momentum of the system and are therefore allowable pulsation
modes is easily extended to the present case, in which we have a solid,
completely rigid core, which is surrounded by a fluid envelope in which
there are pulsations.

In the present case, the center of mass of the fluid in the envelope
{\em is} displaced by the pulsations. However, the pressure variations
associated with these pulsations in the fluid exert a net force on the
crystalline core, causing its center of mass to also move. Considered as
a system, the core plus envelope conserves momentum, so that there is
no net displacement of the center of mass, and $\ell$ = 1 oscillations
are again allowed.


\end{document}